\documentclass[reprint,onecolumn,notitlepage,preprintnumbers,nofootinbib,amsmath,amssymb,aps,prd,floatfix,superscriptaddress,longbibliography,svgnames]{revtex4-1}

\usepackage{amsmath}
\usepackage{amsfonts}
\usepackage{amssymb}
\usepackage{graphicx}
\usepackage{bm}
\usepackage{subfigure}
\usepackage{url}
\usepackage[hyperindex]{hyperref}
\usepackage{color}
\usepackage[ddmmyy,24hr]{datetime}
\usepackage{bigdelim}
\usepackage{booktabs}
\usepackage{dcolumn}
\usepackage{multirow}
\usepackage{subfigure}
\usepackage{physics}
\usepackage{cancel}
\usepackage{stackrel}
\usepackage{paralist}
\usepackage{xspace}
\usepackage{slashed}
\usepackage{cancel}
\usepackage{todonotes}
\usepackage{enumerate}
\usepackage{float}
\usepackage{fullpage}
\usepackage{ulem}
\usepackage{wasysym}
\usepackage{academicons}

\newcommand{\nua}[1]{\ensuremath{\rlap{\kern-2.5pt\ensuremath{\overset{\scriptscriptstyle(-)}{\phantom{\nu}}}}{\ensuremath{{\nu}_{#1}}}}}

\usepackage{enumitem}

\newcommand{\QW}{$Q_{W}$}

\newcommand{\cenns}{CE$\nu$NS\xspace}

\newcommand{\orcid}[1]{\href{https://orcid.org/#1}{\textcolor[HTML]{A6CE39}{\aiOrcid}}}

\begin{document}
\title{Nuclear neutron radius and weak mixing angle measurements from latest COHERENT CsI and atomic parity violation Cs data}

\author{M. Atzori Corona}
\email{mattia.atzori.corona@ca.infn.it}
\affiliation{Dipartimento di Fisica, Universit\`{a} degli Studi di Cagliari,
	Complesso Universitario di Monserrato - S.P. per Sestu Km 0.700,
	09042 Monserrato (Cagliari), Italy}
\affiliation{Istituto Nazionale di Fisica Nucleare (INFN), Sezione di Cagliari,
	Complesso Universitario di Monserrato - S.P. per Sestu Km 0.700,
	09042 Monserrato (Cagliari), Italy}

\author{M. Cadeddu}
\email{matteo.cadeddu@ca.infn.it}
\affiliation{Istituto Nazionale di Fisica Nucleare (INFN), Sezione di Cagliari,
	Complesso Universitario di Monserrato - S.P. per Sestu Km 0.700,
	09042 Monserrato (Cagliari), Italy}

\author{N. Cargioli}
\email{nicola.cargioli@ca.infn.it}
\affiliation{Dipartimento di Fisica, Universit\`{a} degli Studi di Cagliari,
	Complesso Universitario di Monserrato - S.P. per Sestu Km 0.700,
	09042 Monserrato (Cagliari), Italy}
\affiliation{Istituto Nazionale di Fisica Nucleare (INFN), Sezione di Cagliari,
	Complesso Universitario di Monserrato - S.P. per Sestu Km 0.700,
	09042 Monserrato (Cagliari), Italy}

\author{F. Dordei}
\email{francesca.dordei@cern.ch}
\affiliation{Istituto Nazionale di Fisica Nucleare (INFN), Sezione di Cagliari,
 Complesso Universitario di Monserrato - S.P. per Sestu Km 0.700,
 09042 Monserrato (Cagliari), Italy}

\author{C. Giunti}
\email{carlo.giunti@to.infn.it}
\affiliation{Istituto Nazionale di Fisica Nucleare (INFN), Sezione di Torino, Via P. Giuria 1, I--10125 Torino, Italy}

\author{G. Masia}
\email{gbattista98@gmail.com}
\affiliation{Dipartimento di Fisica, Universit\`{a} degli Studi di Cagliari,
 Complesso Universitario di Monserrato - S.P. per Sestu Km 0.700,
	09042 Monserrato (Cagliari), Italy}

\date{\dayofweekname{\day}{\month}{\year} \ddmmyydate\today, \currenttime}

\begin{abstract}
The COHERENT collaboration observed coherent elastic neutrino nucleus scattering using a 14.6 kg cesium-iodide (CsI) detector in 2017 and recently published the updated results before decommissioning the detector. Here, we present the legacy determination of the weak mixing angle and of the average neutron rms radius of $^{133}\mathrm{Cs}$ and $^{127}\mathrm{I}$ obtained with the full CsI dataset, also exploiting the combination with the atomic parity violation (APV) experimental result, that allows us to achieve a precision as low as $\sim$4.5\% and to disentangle the contributions of the $^{133}\mathrm{Cs}$ and $^{127}\mathrm{I}$ nuclei. Interestingly, we show that the COHERENT CsI data show a 6$\sigma$ evidence of the nuclear structure suppression of the full coherence. Moreover, we derive a data-driven APV+COHERENT measurement of the low-energy weak mixing angle with a percent uncertainty, independent of the value of the average neutron rms radius of $^{133}\mathrm{Cs}$ and $^{127}\mathrm{I}$, that is allowed to vary freely in the fit. 
Additionally, we extensively discuss the impact of using two different determinations of the theoretical parity non-conserving amplitude in the APV fit. Our findings show that the particular choice can make a significant difference, up to 6.5\% on $R_n$(Cs) and 11\% on the weak mixing angle. 
Finally, in light of the recent announcement of a future deployment of a 10 kg and a $\sim$700 kg cryogenic CsI detectors, we provide future prospects for these measurements, comparing them with other competitive experiments that are foreseen in the near future.  
\end{abstract}

\maketitle  

\section{Introduction}
\label{sec:intro}

Coherent elastic neutrino-nucleus scattering (\cenns) is a purely weak-neutral-current process, predicted by Freedmann in 1974~\cite{Freedman:1973yd}, which happens at low-momentum transfer when the de Broglie wavelength of the exchanged Z boson mediator is greater than the nucleus radius. In this process, the neutrino interacts with the nucleus as a whole, i.e. coherently, making the cross section roughly proportional to the square of the number of neutrons of the target nucleus. For this reason, it can become some orders of magnitude bigger than that of other low-energy processes which involve neutrinos. Nonetheless, CE$\nu$NS has evaded experimental observation for about 43 years since its first theoretical postulation, due to the difficulty in detecting such a low-energy (few keVs) nuclear recoil produced as the single outcome of the interaction. In fact, the first observation of CE$\nu$NS was reported by the COHERENT collaboration only in 2017~\cite{COHERENT:2017ipa,COHERENT:2018imc}, thanks to 14.6 kg of cesium-iodide (CsI) scintillating crystals exposed to the neutrino flux produced by pion-decays-at-rest at the Spallation Neutron Source (SNS), situated in the Oak Ridge Nation Laboratory. In the subsequent years, the COHERENT collaboration reported a second measurement of \cenns with a liquid argon (LAr) detector~\cite{COHERENT:2020iec,COHERENT:2020ybo}, and later on updated the results with the same CsI detector~\cite{COHERENT:2021xmm}. Furthermore, tantalizing evidence of \cenns using antineutrinos from the Dresden-II boiling water reactor has been recently reported in Ref.~\cite{Colaresi:2022obx} using an ultra-low noise 2.924 kg p-type point-contact germanium detector.\\

In recent years, \cenns has proven to be one of the most profitable tools to study a plethora of diverse physical phenomena, ranging from neutrino non-standard interactions and light mediators~\cite{Liao:2017uzy,Denton:2018xmq,Lindner:2016wff,Flores:2020lji,Coloma:2017ncl,Giunti:2019xpr,Coloma:2022avw,Cadeddu:2020nbr,AtzoriCorona:2022moj,Banerjee:2021laz,Bertuzzo:2021opb} to neutrino electromagnetic properties~\cite{Kosmas:2015sqa,Billard:2018jnl,Coloma:2022avw,Cadeddu:2018dux,AtzoriCorona:2022qrf}, as well as to perform intriguing tests of nuclear physics and electroweak interactions~\cite{Cadeddu:2020lky,Miranda:2020tif,Cadeddu:2021ijh, Cadeddu:2017etk,Papoulias:2019lfi,Papoulias:2017qdn,Cadeddu:2019eta,Papoulias:2019txv,Khan:2019cvi,Dutta:2019eml,AristizabalSierra:2018eqm,Cadeddu:2018izq,Dutta:2019nbn,Abdullah:2018ykz,Ge:2017mcq,Miranda:2021kre,Flores:2020lji,DeRomeri:2022twg}.
Indeed, in previous works~\cite{Cadeddu:2017etk,Cadeddu:2019eta,Papoulias:2019txv,Miranda:2020tif,Huang:2019ene,Cadeddu:2021ijh}, it has been shown that \cenns can be exploited to extract information on the neutron root-mean-square (rms) radius distribution inside atomic nuclei, $R_n$. Such a quantity is of extreme interest in nuclear physics and astrophysics since it provides valuable information on the equation of state (EOS) of neutron-rich matter, which is at the basis of the formation, structure, and stability of atomic nuclei, whether stars collapse into neutron stars or black holes, and also the structure of neutron stars as well as the understanding of gravitational wave events~\cite{Shen:2020sec,Horowitz:2019piw,Wei:2019mdj,Thiel:2019tkm,PREX:2021umo}.
Despite its importance, $R_n$ is experimentally very poorly known and it has been usually determined using strong probes, which are known to be model dependent and affected by non-perturbative effects~\cite{Thiel:2019tkm}. This lack of knowledge has been behind the Lead Radius Experiment (PREX) construction at the Jefferson Laboratory in order to precisely determine $R_n$ of lead, $^{208}$Pb, by measuring the parity-violating electroweak asymmetry in the elastic scattering of polarized electrons from $^{208}$Pb~\cite{Abrahamyan:2012gp,PREX:2021umo}. Notably, the PREX-II collaboration has demonstrated the feasibility of measuring the neutron rms radius of $^{208}$Pb at the percent level~\cite{PREX:2021umo}, being able to obtain the current most accurate determination of the neutron skin, i.e. the difference between $R_n$ and the proton rms radius distribution $R_p$, of a heavy and neutron-rich nucleus like lead.
More recently, the CREX experiment, the twin experiment of PREX that uses calcium, has reported a very precise determination of the neutron radius of ${^{48}\textrm{Ca}}$~\cite{CREX:2022kgg}. These achievements strengthen the importance of further exploiting electroweak probes to obtain direct, reliable and precise determinations of the neutron distribution of neutron-rich nuclei. Luckily, \cenns provides a promising and long-lasting tool, especially as the community is putting a lot of effort into developing the future \cenns program~\cite{Akimov:2022oyb,Baxter:2019mcx,Abdullah:2022zue, NUCLEUS:2019igx,Strauss:2017cuu}.
\\

The \cenns process is also sensitive to the so-called weak mixing angle $\textrm{sin}^2\vartheta_W$, also known as the Weinberg angle, which is a key parameter of the electroweak theory as it describes the mixing of the $\textrm{SU}(2)_L\otimes\textrm{U}(1)_Y$ gauge boson fields~\cite{ParticleDataGroup:2022pth}. Many experiments measured its value very precisely at intermediate or high energies scales~\cite{ParticleDataGroup:2022pth}, while its low-energy value still lacks of a precise experimental determination.
Since the weak mixing angle can be significantly modified in some beyond the standard model (SM) scenarios~\cite{Safronova:2017xyt, Cadeddu:2021dqx},  such as in presence of dark $Z$ bosons ~\cite{Coloma:2022avw,DeRomeri:2022twg,AtzoriCorona:2022moj,Young:2007zs,Cadeddu:2021dqx}, and it may have an impact also for nuclear physics measurements~\cite{Corona:2021yfd}, it is of major importance to exploit the most recent available data to obtain its most accurate state-of-the-art determination using electroweak probes. 
At the moment, the most precise determination of the weak mixing angle available in the low-energy sector comes from another low-energy electroweak probe, so-called Atomic Parity Violation (APV) experiments, also known as parity
non-conservation (PNC) experiments, on cesium atoms~\cite{Wood:1997zq,Guena:2004sq}. In particular, a PNC interaction mixes S and P eigenstates, and, in cesium, this feature permits for the 6S$\rightarrow$7S transition between Zeeman sub-levels~\cite{Gilbert:1985pnq}, that would be otherwise forbidden. In Ref.~\cite{Cadeddu:2018izq,Cadeddu:2021ijh}, it has been demonstrated for the first time the complementarity of \cenns and APV experiments to simultaneously extract information on the weak mixing angle and the nuclear physics parameters, making it interesting to combine these results.\\

The goal of this paper is therefore to use the latest CsI data-release provided by the COHERENT collaboration in 2021 with a refined quenching factor (QF) determination and more statistics~\cite{COHERENT:2021xmm,COHERENT:2021pcd}, both alone and in combination with the information content coming from APV on cesium atoms, in order to extract the most precise information on the weak mixing angle and nuclear physics parameters.
This paper is organized as follows. In Sec.~\ref{sec:theo} we give the necessary theoretical tools for the determination of the physics parameters of interest, while in Sec.~\ref{sec:result} we provide the experimental details and statistical procedures used in this analysis as well as describe and comment the results obtained analysing COHERENT and APV data. Before to summarise our conclusions in Sec.~\ref{sec:conclusions}, in Sec.~\ref{sec:future} we review future perspectives for these measurements.

\section{Nuclear and electroweak physics using \cenns and APV processes}
\label{sec:theo}

As stated in the introduction, \cenns and atomic parity violation represent two of the most promising electroweak probes to access the neutron nuclear distribution.
The proton and neutron nuclear form factors, $F_{Z}\left(|\vec{q}|^{2}\right)$ and $F_{N}\left(|\vec{q}|^{2}\right)$, account respectively for the spatial distribution of protons and neutrons inside the target nucleus. Their measurement can be extracted analysing \cenns data corresponding to a neutrino $\nu_{\ell}$ ($\ell=e,\mu,\tau$) scattering off a nucleus $\mathcal{N}$ with $Z$ protons and $N$ neutrons (in this analysis, we use $(Z,N)_{\mathrm{Cs}}=(55,78)$, and $(Z,N)_{\mathrm{I}}=(53,74)$). They depend on the size of the nucleus and in these kind of processes they parameterize the loss of coherence as the momentum transfer 
$|\vec{q}|$
grows. They represent a crucial ingredient of the \cenns SM 
 differential cross section which, as a function of the nuclear recoil energy $T_\mathrm{nr}$ and for a spin-zero nucleus, is given by~\cite{Freedman:1973yd,Drukier:1984vhf,Barranco:2005yy,Patton:2012jr}
\begin{equation}
	\dfrac{d\sigma_{\nu_{\ell}\text{-}\mathcal{N}}}{d T_\mathrm{nr}}
	(E,T_\mathrm{nr})
	=
	\dfrac{G_{\text{F}}^2 M}{\pi}
	\left( 1 - \dfrac{M T_\mathrm{nr}}{2 E^2} \right)
	\left[g_{V}^{p}\left(\nu_{\ell}\right) Z F_{Z}\left(|\vec{q}|^{2}\right)+g_{V}^{n} N F_{N}\left(|\vec{q}|^{2}\right)\right]^{2},
	\label{eq:cs-std}
\end{equation}
where $G_{\text{F}}$ is the Fermi constant, $E$ is the neutrino energy, $M$ is the nuclear mass, and $g_V^p$ and $g_V^n$ are the neutrino couplings to protons and neutrons, respectively. Their tree-level values are $g_V^p=1/2-2\;\textrm{sin}^2\;\vartheta_{\text{W}}$, which encapsulates the dependence on the weak mixing angle such that it could be determined from \cenns data, and $g_V^n=-1/2$. In the SM, the value of the weak mixing angle $\vartheta_W$ evaluated at zero-momentum transfer is equal to $\textrm{sin}^2\vartheta_{\text{W}}(q^2=0)=\hat{s}^2_0=0.23863$, as reported in the Particle Data Group (PDG)~\cite{ParticleDataGroup:2022pth}. 
A precise determination of these couplings is of fundamental importance to retrieve correctly $F_{Z}\left(|\vec{q}|^{2}\right)$ and $F_{N}\left(|\vec{q}|^{2}\right)$. Thus, in this paper, we calculated the couplings taking into account the radiative corrections in the $\overline{\mathrm{MS}}$ scheme following Refs.~\cite{Erler:2013xha,ParticleDataGroup:2022pth}
\begin{equation}
\!
\begin{aligned}
g_V^{\nu_\ell\,p}&=\rho\left(\frac{1}{2}-2\;\textrm{sin}^2\vartheta_{\text{W}}\right)+2\; {\raisebox{-1pt}{\rotatebox{90}{\Bowtie}}}_{WW}+\Box_{WW}-2\diameter_{\nu_\ell W}+\rho(2\boxtimes_{ZZ}^{uL}+\boxtimes_{ZZ}^{dL}-2\boxtimes_{ZZ}^{uR}-\boxtimes_{ZZ}^{dR})\,, \\
g_V^{\nu_\ell\,n}&=-\frac{\rho}{2}+2\Box_{WW}+{\raisebox{-1pt}{\rotatebox{90}{\Bowtie}}}_{WW}+\rho(2\boxtimes_{ZZ}^{dL}+\boxtimes_{ZZ}^{uL}-2\boxtimes_{ZZ}^{dR}-\boxtimes_{ZZ}^{uR}).
\end{aligned}\label{eq:gv}
\end{equation}

The quantities in Eq.~(\ref{eq:gv}), $\Box_{WW}$, ${\raisebox{-1pt}{\rotatebox{90}{\Bowtie}}}_{WW}$ and $\boxtimes_{ZZ}^{fX}$, with $f\in\{u,d\}$ and $X\in\{L,R\},$ are the radiative corrections associated with the $WW$ box diagram, the $WW$ crossed-box and the $ZZ$ box respectively, while $\rho=1.00063$ is a parameter of electroweak interactions. Moreover, $\diameter_{\nu_\ell W} $ describes the neutrino charge radius contribution and introduces a dependence on the neutrino flavour $\ell$ (see Ref.~\cite{Erler:2013xha} or the appendix B of Ref.~\cite{AtzoriCorona:2022jeb} for further information on such quantities).
Numerically, the values of these couplings correspond to $g_{V}^{p}(\nu_{e}) = 0.0382$, $g_{V}^{p}(\nu_{\mu}) = 0.0300$, and $g_{V}^{n} = -0.5117$. \\

The form factors $F_{N}\left(|\vec{q}|^{2}\right)$ and $F_{Z}\left(|\vec{q}|^{2}\right)$ can be parametrized using a symmetrized Fermi~\cite{Cadeddu:2017etk} distribution which is given by the analytic expression
\begin{equation}
F_{Z}^{\text{SF}}(q^2)
=
\dfrac{ 3 }{ q c \left[ ( q c )^2 + ( \pi q a )^2 \right] }
\left[
\dfrac{ \pi q a }{ \sinh( \pi q a ) }
\right]
\left[
\dfrac{ \pi q a \sin( q c ) }{ \tanh( \pi q a ) }
-
q c \cos( q c )
\right]
.
\label{ffzSF}
\end{equation}
The parameter $a$ is the so-called diffuseness, which is related to the surface thickness $t$ through $t = 4\;a\;\textrm{ln}(3)$ and is commonly fixed to the value $t=2.30$ fm both for the neutron and proton form factor, as most of theoretical nuclear models predict roughly the same density drop between proton and neutron distributions. The rms radius is related to $a$ through $R^2 = 3/5 \, c^2 + 7/5 \, ( \pi a )^2$ and, for the proton distribution inside the nucleus, we use the proton rms radii obtained from the muonic atom spectroscopy data~\cite{Fricke:1995zz,Angeli:2013epw} as explained in Ref.~\cite{Cadeddu:2020lky}, namely $R_{p}(\mathrm{Cs})=4.821(5)~\mathrm{fm}$ and $R_{p}(\mathrm{I})=4.766(8) ~\mathrm{fm}$. Since only little knowledge on the neutron rms radii of ${}^{133}\text{Cs}$ and ${}^{127}\text{I}$ is available using electroweak probes~\cite{Hoferichter:2020osn,Cadeddu:2017etk,Papoulias:2019lfi,Cadeddu:2018dux,Huang:2019ene,Papoulias:2019txv,Khan:2019cvi,Cadeddu:2020lky,Cadeddu:2019eta}, we take as a reference the values of the neutron rms radii extracted from the theoretical nuclear shell model (NSM) calculations of the corresponding neutron skins $\Delta R_{\rm{np}}$~\cite{Hoferichter:2020osn} which are
\begin{align}\label{Rn}
	R^{\rm NSM}_{n}(^{133}\mathrm{Cs})\simeq5.09~\mathrm{fm}, \quad\quad R^{\rm NSM}_{n}(^{127}\mathrm{I})\simeq5.03~\mathrm{fm}.
\end{align}
In the calculations which assume these values of the neutron rms radii,
we take into account the effect of their uncertainties by considering a 3.4\% uncertainty for the CsI \cenns rates~\cite{COHERENT:2021xmm}.

An alternative description is provided by the Helm parameterization~\cite{Helm:1956zz}, which practically gives the same results as the SF.
The Helm parameterization is given by
\begin{equation}\label{eq:Helm} F^\textrm{Helm}\left(|\vec{q}|^{2}\right)=3\frac{j_1(qR_0)}{qR_0}e^{-|\vec{q}|^{2}s^2/2},
\end{equation}
where $j_1(x)=\textrm{sin}(x)/x^2-\textrm{cos}(x)$ is the order-one spherical Bessel function, while $R_0$ is the box (or diffraction) radius.
The rms radius of the corresponding nucleon distribution is given by $R^2=3/5 R_0^2+3s^2$,
where the parameter $s$ quantifies the so-called surface thickness.
We consider a value of $s=0.9$ fm, which is the typical value determined for the proton form factor for this type of nuclei~\cite{Friedrich:1982esq}. \\

The low-energy measurement of the nuclear weak charge, $Q_W$, of $^{133}\text{Cs}$ from APV experiments also depends on the nuclear physics parameters and the weak mixing angle, making it interesting to simultaneously constrain these parameters exploiting COHERENT CsI data.
In particular, the nuclear weak charge depends on the weak mixing angle according to the relation~\cite{ParticleDataGroup:2022pth}
\begin{align}
Q_W^{\mathrm{th}}(\sin^2\vartheta_W)
=
&
- 2 [ Z (g_{A V}^{e p}(\sin^2\vartheta_W) + 0.00005) + N (g_{A V}^{e n} + 0.00006) ] 
\left( 1 - \dfrac{\alpha}{2 \pi} \right)
,
\label{QWSMthe}
\end{align}
where $\alpha$ is the fine-structure constant, while $g_{A V}^{e p}$ and $g_{A V}^{e n}$ are the couplings of electrons to protons and neutrons, respectively, including radiative corrections in the $\overline{\mathrm{MS}}$ scheme~\cite{ParticleDataGroup:2022pth,Erler:2013xha}
(see also appendix A of Ref.~\cite{Cadeddu:2021ijh}). The small numerical corrections to the couplings are discussed in Ref.~\cite{Erler:2013xha} and include the calculation of the $\gamma Z$-box correction from Ref.~\cite{Blunden:2012ty}.
The values of the couplings correspond to
\begin{align}
\null & \null
g_{A V,\text{SM}}^{e p}
=
2 g_{A V,\text{SM}}^{e u} + g_{A V,\text{SM}}^{e d}
=
- 0.0357
,
\label{gAVeppdg}
\\
\null & \null
g_{A V,\text{SM}}^{e n}
=
g_{A V,\text{SM}}^{e u} + 2 g_{A V,\text{SM}}^{e d}
=
0.495
,
\label{gAVenpdg}
\end{align}
where $g_{A V,\text{SM}}^{e u}
=
- 0.1888$ and $g_{A V,\text{SM}}^{e d}
=
0.3419$. These values give
$Q_{{W}}^{\text{SM}}
=
-73.23 \pm 0.01
\label{QWSMval}$.\\

Experimentally, the weak charge of a nucleus is extracted from the ratio of the parity violating amplitude, $E_{\text{PNC}}$, to the Stark vector transition polarizability, $\beta$, and by calculating theoretically $E_{\rm PNC}$ in terms of \QW
\begin{equation}
\label{QWeq}
Q_W= N \left( \dfrac{{\rm Im}\, E_{\rm PNC}}{\beta} \right)_{\rm exp.} 
\left( \dfrac{Q_W}{N\, {\rm Im}\, E_{\rm PNC}} \right)_{\rm th.} \beta_{\rm exp.+th.}\,,
\end{equation}

where $\beta_{\rm exp.+th.}$ and $(\mathrm{Im}\, E_{\rm PNC})_{\rm th.}$ are determined from atomic theory, and Im stands for imaginary part~\cite{ParticleDataGroup:2022pth}.
We use $({\rm Im}\, E_{\rm PNC}/{\beta})_{\rm exp} = (-3.0967 \pm 0.0107) \times 10^{-13} |e|/a_B^2$~\cite{ParticleDataGroup:2022pth}, where $a_B$ is the Bohr radius and $|e|$ is the absolute value of the electric charge, and $\beta_{\rm exp.+th.} = (27.064 \pm 0.033)\, a_B^3$~\cite{ParticleDataGroup:2022pth}.\\
In order to extract information on the neutron distribution inside cesium and to directly evaluate $R_n$ from a combined fit with the COHERENT data, one has to subtract to $({\rm Im}\, E_{\rm PNC})_{\rm th.}$ the so-called ``neutron skin" correction $\delta E^\mathrm{n.s.}_\mathrm{PNC}(R_n)$ in order to obtain $({\rm Im}\, E_{\rm PNC})_{\rm th.}^{\rm w.n.s.}$~\cite{Cadeddu:2021ijh}.
This correction has been introduced in Ref.~\cite{PhysRevA.65.012106} to take into account the fact that the difference between $R_n$ and $R_p$ is not considered in the nominal atomic theory derivation. Hence, the neutron skin corrected value of the weak charge, $Q_W^\mathrm{n.s.}(R_n)$, is retrieved by summing to $({\rm Im}\, E_{\rm PNC})_{\rm th.}^{\rm w.n.s.}$ the correction term $\delta E^\mathrm{n.s.}_\mathrm{PNC}(R_n)= \left[ (\mathrm{N}/Q_W)\left(1-(q_n(R_n)/q_p)\right) E_\mathrm{PNC}^\mathrm{w.n.s.} \right]$~\cite{PhysRevA.65.012106,Cadeddu:2019eta,Cadeddu:2018izq}. 
In this way, the dependence of $Q_W$ on the neutron rms radius becomes also explicit. Indeed, the factors $q_p$ and $q_n$ incorporate the radial dependence of the electron axial transition matrix element by considering the proton and the neutron spatial distribution, respectively~\cite{PhysRevC.46.2587,PhysRevC.100.034318,PhysRevC.46.2587,PhysRevA.65.012106}.
For the calculation of $q_p$ and $q_n$ we refer to appendix B of Ref.~\cite{Cadeddu:2021ijh}.
In this work, we will rely on the theoretical value of the PNC amplitude with the neutron skin correction used in Ref.~\cite{Cadeddu:2021ijh} $({\rm Im}\, E_{\rm PNC})_{\rm th.}^{\rm w.n.s.}=(0.8995\pm0.0040)\times10^{-11}|e|a_B \frac{Q_W}{N}$~\cite{Dzuba:2012kx}, but we will also discuss the implications to consider the result of the more recent calculation reported in Ref.~\cite{Sahoo:2021thl} which yields to a smaller value, namely $({\rm Im}\, E_{\rm PNC})_{\rm th.}^{\rm w.n.s.}=(0.8930\pm0.0027)\times10^{-11}|e|a_B \frac{Q_W}{N}$.

\section{Results}
\label{sec:result}

In this section, we present the new measurements of the weak mixing angle, performed at the energy scale probed by the experiments, i.e. $|\vec{q}|^2\simeq (50\  {\rm{MeV}})^2$ and $|\vec{q}|^2\simeq (2.4\  {\rm{MeV}})^2$ for COHERENT and APV, respectively, and of the cesium nuclear parameters, both obtained using the latest COHERENT CsI data-set~\cite{COHERENT:2021xmm} alone and in combination with APV.\\

Concerning the treatment of the COHERENT data, we largely follow the same procedure described in details in Ref.~\cite{AtzoriCorona:2022moj}, to which we refer the reader. The physical parameters of interest are extracted using the following Poissonian least-squares function~\cite{Baker:1983tu,ParticleDataGroup:2022pth}
\begin{align}
	\chi^2_{\mathrm{CsI}}
	=
	\null & \null
	2
	\sum_{i=1}^{9}
	\sum_{j=1}^{11}
	\left[
	    \sum_{z=1}^{4}( 1 + \eta_{z} ) N_{ij}^{z} -
		N_{ij}^{\text{exp}}
		+ N_{ij}^{\text{exp}} \ln\left(\frac{N_{ij}^{\text{exp}}}{\sum_{z=1}^{4}( 1 + \eta_{z} ) N_{ij}^{z}}\right)
	\right]
	+ \sum_{z=1}^{4}
	\left(
	\dfrac{ \eta_{z} }{ \sigma_{z} }
	\right)^2
	,
	\label{chi2coherentCsI}
\end{align}
where
the indices $i,j$ represent the nuclear-recoil energy and arrival time bin, respectively, while the indices 
$z=1,2,3,4$ for $N_{ij}^{z}$ stand, respectively, for CE$\nu$NS, ($N_{ij}^{1}=N_{ij}^{\mathrm{CE}\nu\mathrm{NS}}$), beam-related neutron ($N_{ij}^{2}=N_{ij}^{\text{BRN}}$), neutrino-induced neutron ($N_{ij}^{3}=N_{ij}^{\text{NIN}}$) and steady-state ($N_{ij}^{4}=N_{ij}^{\text{SS}}$) backgrounds obtained from the anti-coincidence data.
In our notation, $N_{ij}^{\text{exp}}$ is the experimental event number obtained from coincidence data and $N_{ij}^{\text{\cenns}}$ is the predicted number of \cenns events that depends on the physics model under consideration, according to the cross-section in Eq.~(\ref{eq:cs-std}), as well as on the neutrino flux, energy resolution, detector efficiency, number of target atoms and the CsI quenching factor~\cite{AtzoriCorona:2022moj}. 
We take into account the systematic uncertainties with the nuisance parameters $\eta_{z}$
and the corresponding uncertainties
$\sigma_{\text{CE}\nu\text{NS}}=0.12$,
$\sigma_{\text{BRN}}=0.25$,
$\sigma_{\text{NIN}}=0.35$ and
$\sigma_{\text{SS}}=0.021$ as explained in Refs.~\cite{AtzoriCorona:2022moj,COHERENT:2021xmm}. \\

When performing the analysis of the APV data, we use the least-squares function given by
\begin{align}
	\chi^2_{\mathrm{APV}} 	=
    \left(
    \dfrac{
    Q_W^{\rm Cs\,n.s.}(R_n)
    -
    Q_W^{\mathrm{th}}(\sin^2\vartheta_W)
    }{ \sigma_{\rm APV}(R_n,\sin^2\vartheta_W) }
    \right)^2
    \,
    ,
    \label{chiapvRn}
\end{align}
where $\sigma_{\rm APV}$ is the total uncertainty.
Finally, when performing a combined analysis of the COHERENT data with APV, we use the least-squares function given by
\begin{align}
	\chi^2 	=
	\chi^2_{\mathrm{CsI}}
    +\chi^2_{\mathrm{APV}}.
\label{chicoherentapvRn}
\end{align}

\subsection{Determination of $\mathbf{\sin^2\vartheta_W}$ and the average CsI neutron radius}\label{sec:1dsin2}

\begin{figure*}[!t]
\centering
\setlength{\tabcolsep}{0pt}
\begin{tabular}{cc}
\subfigure[]{\label{fig:1Dsin}
\begin{tabular}{c}
\includegraphics*[width=0.42\linewidth]{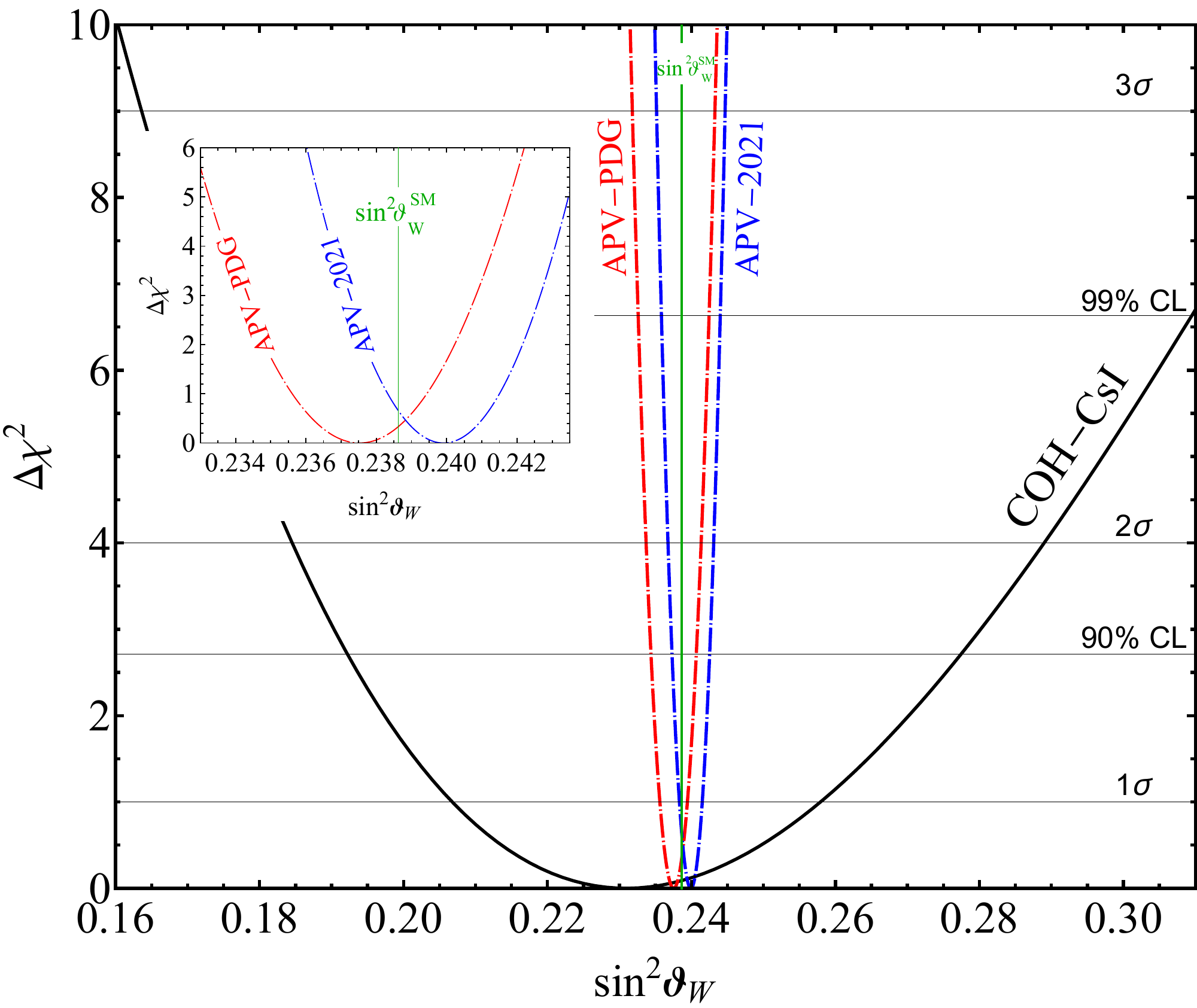}
\\
\end{tabular}
}
&
\subfigure[]{\label{fig:1DRn}
\begin{tabular}{c}
\includegraphics*[width=0.42\linewidth]{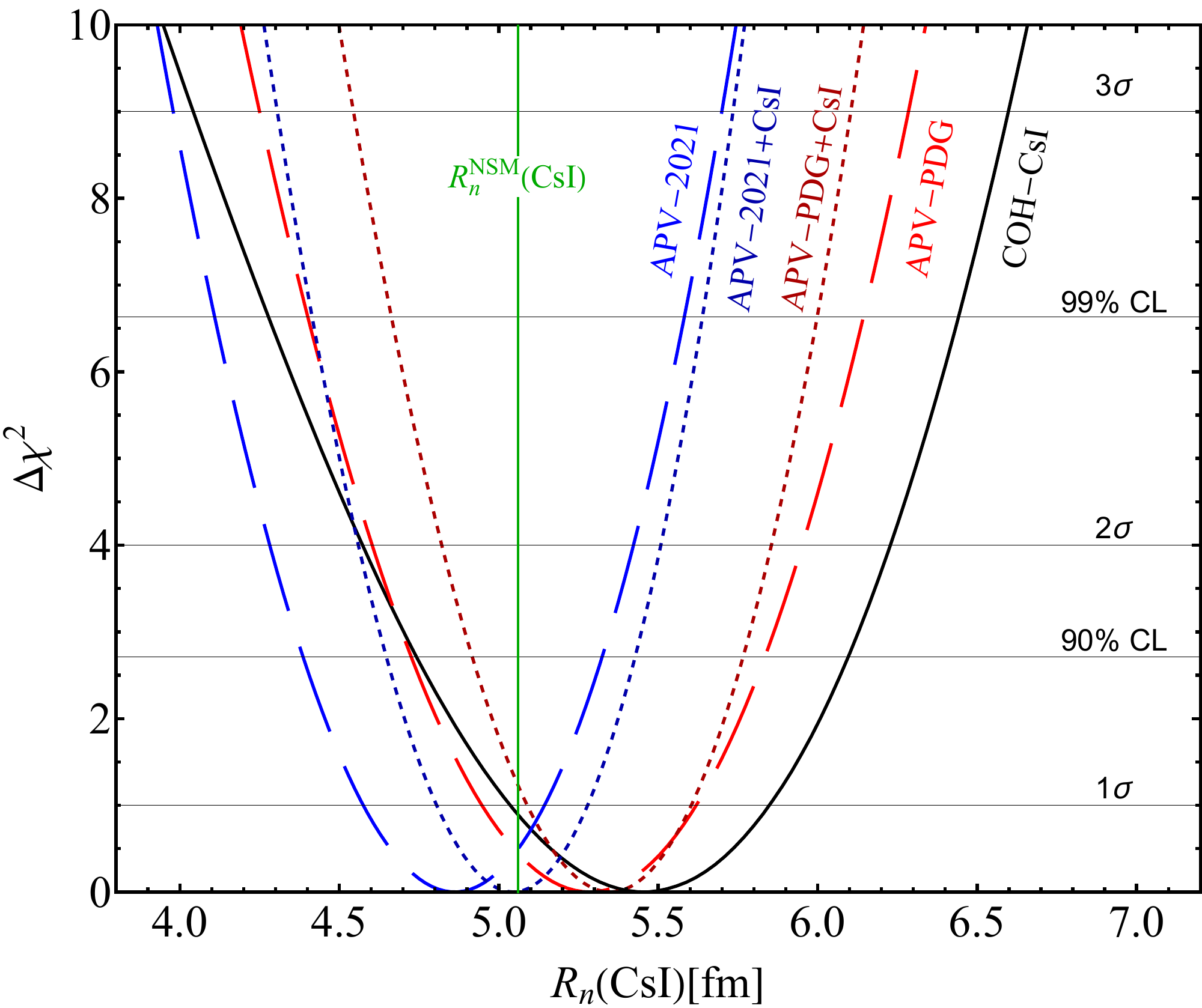}
\\
\end{tabular}
}
\\
\end{tabular}
\caption{ \label{fig:1D}
Constraints on the weak mixing angle (a) and on the average rms CsI neutron radius (b) at different confidence levels (CL). The different curves refer to the COHERENT CsI data (CsI), the APV data using the PNC amplitude of Ref.~\cite{Dzuba:2012kx} (APV-PDG) and that recently calculated in Ref.~\cite{Sahoo:2021thl} (APV-2021), as well as their combination (APV-PDG+CsI and APV-2021+CsI). In (a) the combined curves are practically indistinguishable from the APV only fit. The green lines represent (a) the low-energy SM value of the weak mixing angle and  (b) the average rms CsI neutron radius from the nuclear shell model prediction in Eq.~\eqref{Rn}. In the inset in the top left of (a), a zoom of x-axes is shown to better appreciate the APV only determinations.
}
\end{figure*}

\begin{table}
{
{\renewcommand{\arraystretch}{1.2} 
\begin{tabular}{l|c|c|c|c|}
& \multicolumn{2}{c|}{$\sin^2\vartheta_W$} & \multicolumn{2}{c}{$R_n(\rm{CsI}) [fm]$}\\ 
& \multicolumn{1}{c|}{$\textrm{best-fit}^{+1\sigma+90\%\rm{CL}+2\sigma}_{-1\sigma-90\%\rm{CL}-2\sigma}$} & $\chi^2_{\rm{min}}$ & $\textrm{best-fit}^{+1\sigma+90\%\rm{CL}+2\sigma}_{-1\sigma-90\%\rm{CL}-2\sigma}$ & $\chi^2_{\rm{min}}$

\\[1mm]\hline
COH-CsI & $0.231^{+0.027+0.046+0.058}_{-0.024-0.039-0.047} $ & 86.0 & $5.47^{+0.38+0.63+0.76}_{-0.38-0.72-0.89}$ & 85.2\\[0.5mm] 
APV~PDG & $0.2375^{+0.0019+0.0031+0.0038}_{-0.0019-0.0031-0.0038}$ & - & $5.29^{+0.33+0.
55+0.66}_{-0.34-0.56-0.68}$ & -\\[0.5mm]
APV~2021 & $0.2399^{+0.0016+0.0026+0.0032}_{-0.0016-0.0026-0.0032}$ & - & $4.86^{+0.28+0.
46+0.56}_{-0.29-0.48-0.58}$ & -\\[0.5mm]
APV~PDG + CsI & $0.2374^{+0.0020+0.0032+0.0039}_{-0.0018-0.0031-0.0037}$ & 86.0 & $5.35^{+0.25+0.41+0.50}_{-0.26-0.43-0.53}$ & 85.3\\[0.5mm]
APV~2021 + CsI & $0.2398^{+0.0016+0.0026+0.0032}_{-0.0015-0.0026-0.0031}$ & 86.0 & $5.04^{+0.23+0.38+0.46}_{-0.24-0.40-0.48}$ & 86.6\\
\end{tabular}

%

}
}
\caption{Summary of the constraints obtained in this work on the weak mixing angle $\sin^2{\vartheta_{\text{W}}}$ and on the average rms CsI neutron radius $R_{n}(\rm{CsI})$ at different confidence levels. The different labels refer to the COHERENT CsI data (COH- CsI), the APV data using the PNC amplitudes of Ref.~\cite{Dzuba:2012kx} (APV-PDG) and Ref.~\cite{Sahoo:2021thl} (APV-2021), as well as their combination (APV-PDG+CsI and APV-2021+CsI). For each constrain we also report the minimum value of the least square function provided by the fit.}  \label{tab:tablimits}
\end{table}

In this section, we report the limits obtained on $\sin^2{\vartheta_{\text{W}}}$ and $R_{n}(\rm{CsI})$ separately, i.e., by fitting for only one parameter at a time and fixing the other to the theoretical prediction. To this end, we exploit the COHERENT CsI and APV dataset separately but also their combination. \\

In Sec.~\ref{sec:intro} we introduced the weak mixing angle $\vartheta_{\text{W}}$, whose experimental determination might provide a powerful tool to test the SM of electroweak theory and to investigate possible new physics extensions of it. 
With this aim in mind, we updated the previous COHERENT CsI analysis performed in Ref.~\cite{Cadeddu:2021ijh} using a refined quenching factor~\cite{COHERENT:2021xmm,COHERENT:2021pcd} and an improved fitting procedure which exploits the neutrino time arrival information and uses a poissonian $\chi^2$ function, instead of the canonical gaussian definition, which provides a more correct and reliable fit to the data when the number of events in each bin is small.
Relying on the theoretical prediction for the neutron radius, we can fix the average CsI radius to $R_{n}^{\rm{NSM}}(\rm{CsI})\simeq 5.06\; \rm{fm}$, according to Eq.~(\ref{Rn}), to perform a fit to extract the weak mixing angle using the least-square function in Eq.~(\ref{chi2coherentCsI}). The result is shown graphically in Fig.~\ref{fig:1Dsin} and at the $1\sigma$, 90$\%$ and $2\sigma$ confidence level (CL) we find
\begin{equation}
  \sin^2{\vartheta_{\text{W}}}(\mathrm{COH-CsI}) = 0.231^{+0.027}_{-0.024} (1\sigma)^{+0.046}_{-0.039}(90\% \textrm{CL})^{+0.058}_{-0.047} (2\sigma) ,
\end{equation}
which is in agreement with the theoretical SM prediction $\hat{s}^2_0$ and the result recently presented in Ref.~\cite{DeRomeri:2022twg} when fitting the COHERENT CsI data with a different approach. Another derivation performed by the COHERENT collaboration~\cite{COHERENT:2021xmm} reports $\sin^2{\vartheta_{\text{W}}}(\mathrm{CsI}) = 0.220^{+0.028}_{-0.026}$, which agrees rather well with our result although some small differences are expected due to the different description of the nuclear structure, i.e. different choices of the reference values for the neutron nuclear radii, and a different approach to radiative corrections for neutrino-nucleus scattering. Moreover, we checked the impact of using a different quenching factor, by comparing our nominal results obtained using Refs.~\cite{COHERENT:2021xmm,COHERENT:2021pcd} and the derivation in Ref.~\cite{Lewis:2021cjv}. The latter lower QF decreases the total number of \cenns events resulting in a larger $\sin^2{\vartheta_{\text{W}}}$ by about 10\%.

This result can be compared with those obtained using the APV experiment on Cs and the least-squares function in Eq.~(\ref{chiapvRn}), namely
\begin{eqnarray}
  \sin^2{\vartheta_{\text{W}}}(\mathrm{APV\,PDG}) = 0.2375\pm0.0019\, (1\sigma)\pm0.0031\,(90\% \textrm{CL})\pm0.0038\, (2\sigma) ,\\
    \sin^2{\vartheta_{\text{W}}}(\mathrm{APV\,2021}) = 0.2399\pm0.0016\, (1\sigma)\pm0.0026\,(90\% \textrm{CL})\pm0.0032\, (2\sigma),
\end{eqnarray}
that we derived exploiting the experimental value of $Q_W$ obtained with the theoretical prediction of the PNC amplitude of Ref.~\cite{Dzuba:2012kx}, referred to as APV PDG, and that recently calculated in Ref.~\cite{Sahoo:2021thl}, referred to as APV 2021. It is possible to see that the APV dataset allows us to achieve a factor of more than 10 better precision in the determination of $\sin^2{\vartheta_{\text{W}}}$. As shown in the insert of  Fig.~\ref{fig:1Dsin}, the two PNC amplitudes point to a value of the weak mixing angle that is below and above the theoretical prediction, respectively, by less than 1$\sigma$. From the combination of the APV and COHERENT CsI dataset we obtain
\begin{eqnarray}
  \sin^2{\vartheta_{\text{W}}}(\mathrm{APV\,PDG+COH-CsI}) = 0.2374^{+0.0020}_{-0.0018} (1\sigma)^{+0.0032}_{-0.0031}(90\% \textrm{CL})^{+0.0039}_{-0.0037} (2\sigma) ,\\
    \sin^2{\vartheta_{\text{W}}}(\mathrm{APV\,2021+COH-CsI}) = 0.2398^{+0.0016}_{-0.0015} (1\sigma)^{+0.0026}_{-0.0026}(90\% \textrm{CL})^{+0.0032}_{-0.0031} (2\sigma).
\end{eqnarray}
Clearly, the combination is vastly dominated by the APV result, with the PNC amplitude from Ref.~\cite{Sahoo:2021thl} being slightly more precise. All the results shown are summarised in Table~\ref{tab:tablimits}.\\

\begin{figure}
    \centering
    \includegraphics[width=0.8\textwidth]{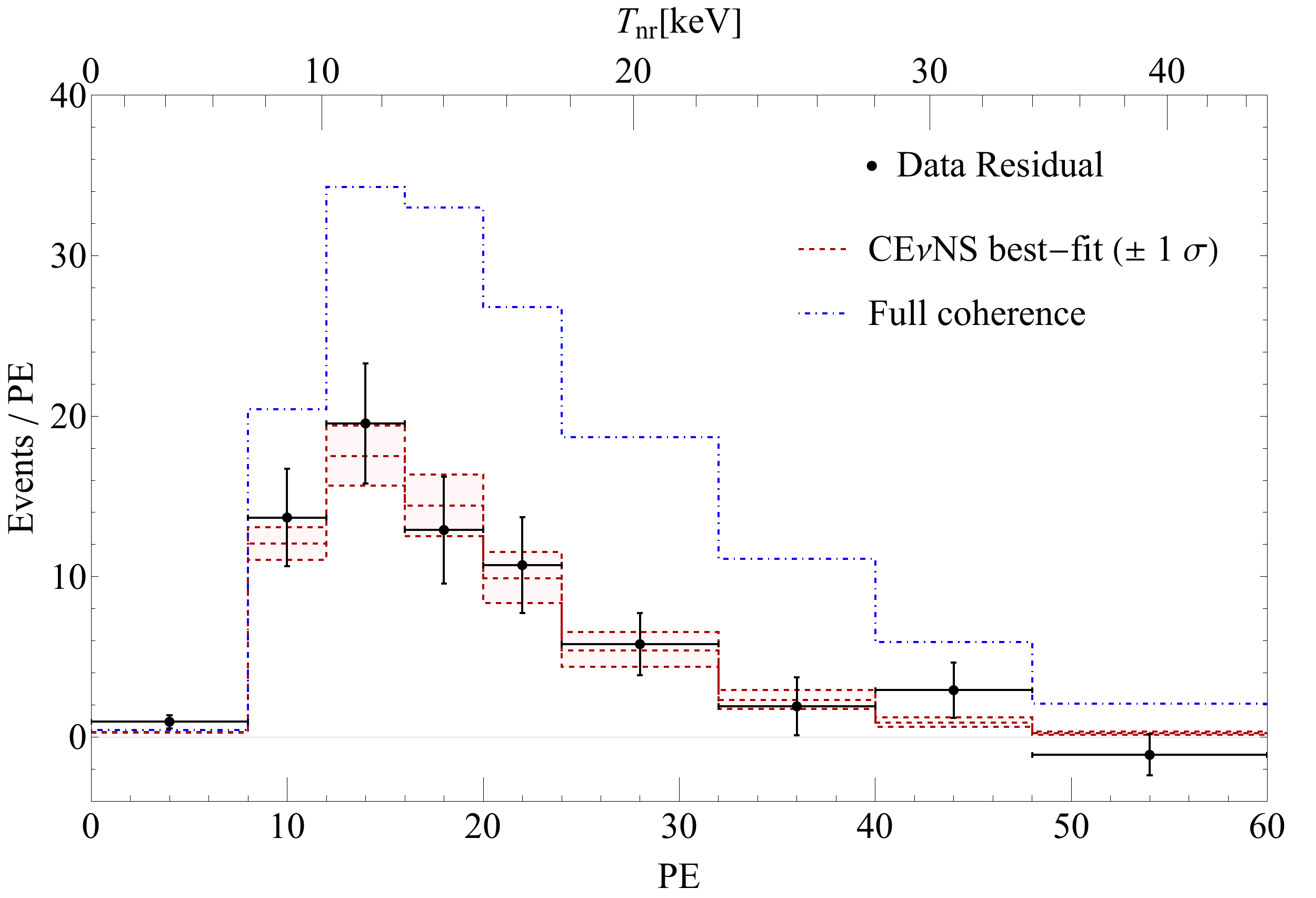}
    \caption{COHERENT \cenns only data versus the number of photoelectrons (PE) and the nuclear kinetic recoil energy ($T_{\mathrm{nr}}$). The histograms represent the theoretical prediction in the case of full coherence (blue dash-dotted line) and the best fit obtained leaving $R_n$ free to vary (red dashed line). The red shadowed area represents the $\pm1\sigma$ variation in the $R_n$ value.}
    \label{fig:Coherence}
\end{figure}

Moving now to the poorly-known neutron distribution inside the nuclei, in order to obtain information on it we fix the weak mixing angle to the SM low-energy value, that is currently calculated with very high precision~\cite{ParticleDataGroup:2022pth}, to let the average CsI neutron distribution radius $R_{n}(\rm{CsI})$ free to vary in the fit. Indeed, given the fact that the difference between the rms neutron radii of Cs and I is expected to be small compared to the current precision of experimental data, the choice to fit for an average value is a fair approximation. 
Clearly, in this case, the contribution due to the neutron form factor to the total systematic uncertainty on $N_{ij}^{\mathrm{CE}\nu\mathrm{NS}}$ is removed in the least-square function evaluation. The result of the fit is shown in Fig.~\ref{fig:1DRn} and at the $1\sigma$, 90$\%$ and $2\sigma$ CL we find
\begin{equation}
  R_n(\mathrm{COH-CsI}) = 5.47^{+0.38}_{-0.38} (1\sigma)^{+0.63}_{-0.72}(90\% \textrm{CL})^{+0.76}_{-0.89} (2\sigma)\;{\rm{fm}},
\end{equation}
which is in agreement, within the uncertainty, with the NSM expected value for $R_{n}^{\rm{NSM}}(\rm{CsI})$, despite the central value pointing toward a large neutron skin. Moreover, this result is almost $10\%$ more precise than the previous determination of Ref.~\cite{Cadeddu:2021ijh}. 
To better appreciate the sensitivity of \cenns to $R_n$, in Fig.~\ref{fig:Coherence} we show the impact of the nuclear structure to the theoretical prediction of the \cenns event rates. In particular, we show the COHERENT excess counts, namely the background subtracted COHERENT data,
as a function of both the photoelectrons (PE) and the corresponding nuclear recoil energy ($T_{\rm{nr}}$) and we compare them
with the prediction obtained in case of full coherence, i.e., setting all nuclear form factors equal to unity, and with the best fit obtained leaving $R_n$ free to vary, as described in this section.  We find that COHERENT data shows a 6$\sigma$ evidence of the nuclear structure suppression of the full coherence, making it an extremely powerful probe to determine nuclear parameters.

Also APV data is sensitive to $R_n$ and using the least-squares function in Eq.~(\ref{chiapvRn}) we get
\begin{eqnarray}
  R_n(\mathrm{\mathrm{APV\,PDG}}) = 5.29^{+0.33}_{-0.34} (1\sigma)^{+0.55}_{-0.56}(90\% \textrm{CL})^{+0.66}_{-0.68} (2\sigma)\;{\rm{fm}},\\
    R_n(\mathrm{\mathrm{APV\,2021}}) = 4.86^{+0.28}_{-0.29} (1\sigma)^{+0.46}_{-0.48}(90\% \textrm{CL})^{+0.56}_{-0.58} (2\sigma)\;{\rm{fm}}.
\end{eqnarray}
Differently from the case of the weak mixing angle, the precision achieved in this case is only slightly better than that achieved with COHERENT, such that the constraints improve significantly by performing a combination of these two experiments. The $\chi^2$-curves we obtain are summarised in Fig.~\ref{fig:1DRn} and the numerical values we find are
\begin{eqnarray}
  R_n(\mathrm{\mathrm{APV\,PDG+COH-CsI}}) = 5.35^{+0.25}_{-0.26} (1\sigma)^{+0.41}_{-0.43}(90\% \textrm{CL})^{+0.50}_{-0.53} (2\sigma)\;{\rm{fm}},\\
    R_n(\mathrm{\mathrm{APV\,2021+COH-CsI}}) = 5.04^{+0.23}_{-0.24} (1\sigma)^{+0.38}_{-0.40}(90\% \textrm{CL})^{+0.46}_{-0.48} (2\sigma)\;{\rm{fm}}.
\end{eqnarray}
It is possible to see that the combination obtained using the 2021 PNC amplitude of Ref.~\cite{Sahoo:2021thl} returns a neutron distribution rms radius that is very well in agreement with the theoretical prediction, while both COHERENT and 
the PDG PNC amplitude of Ref.~\cite{Dzuba:2012kx} suggest a larger neutron skin. Given that in the latter case the two dataset point toward a similar value, we also get a smaller value for the minimum $\chi^2$, as shown in Table~\ref{tab:tablimits}. In both scenarios, a precision of less than 5\% is obtained in the determination of $R_n$.

\subsection{Simultaneous determination of $\mathbf{\sin^2\vartheta_W}$ vs $\mathbf{R_{n}}$(CsI) and combined analysis with APV}\label{sec:Csi2DRnsin2}

\begin{figure*}[!]
\centering
\setlength{\tabcolsep}{0pt}
\begin{tabular}{cc}
\subfigure[]{\label{fig:2Dsin2rna}
\begin{tabular}{c}
\includegraphics*[width=0.42\linewidth]{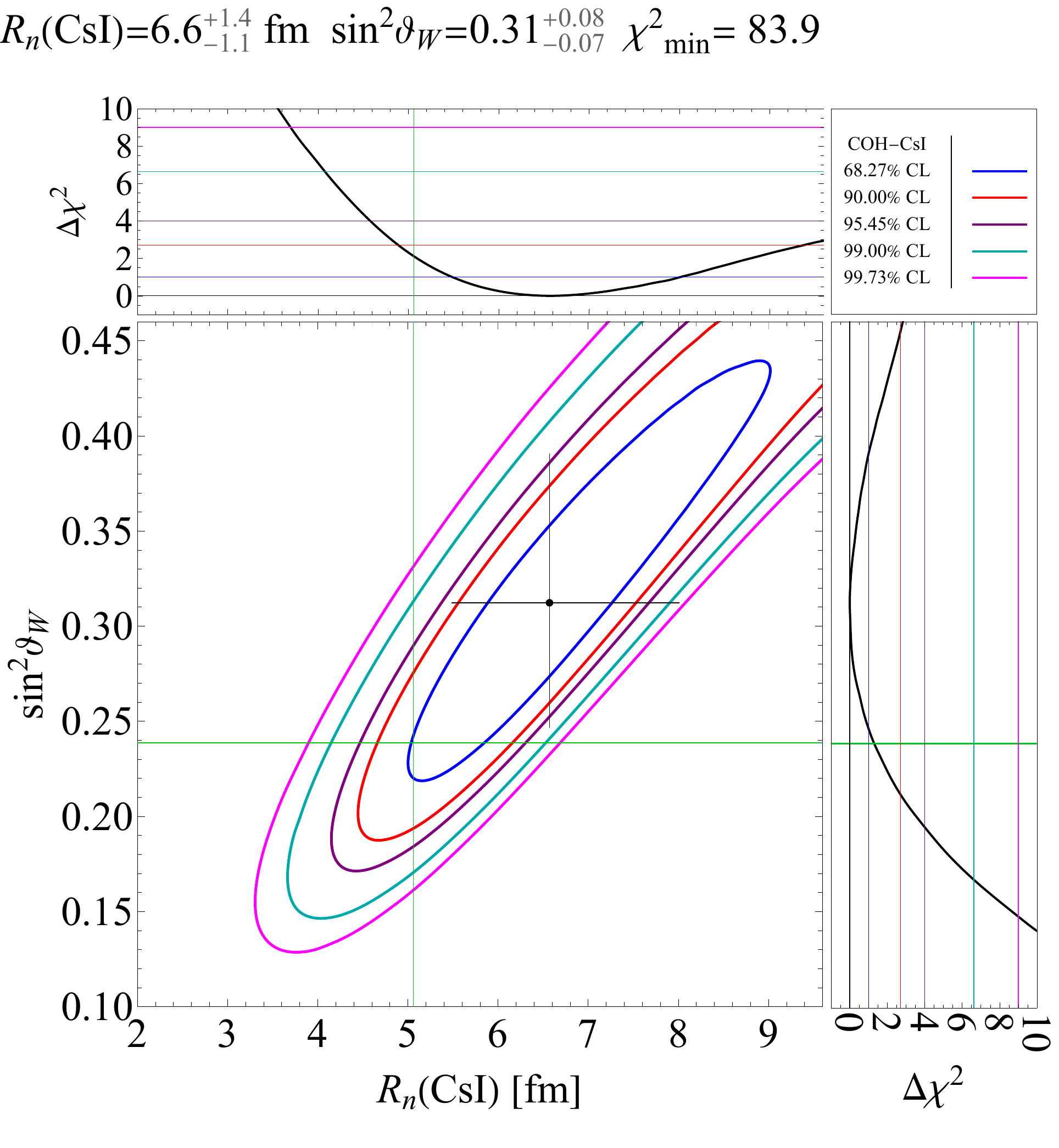}
\\
\end{tabular}
}
&
\subfigure[]{\label{fig:2DrnCsrnI}
\begin{tabular}{c}
\includegraphics*[width=0.42\linewidth]{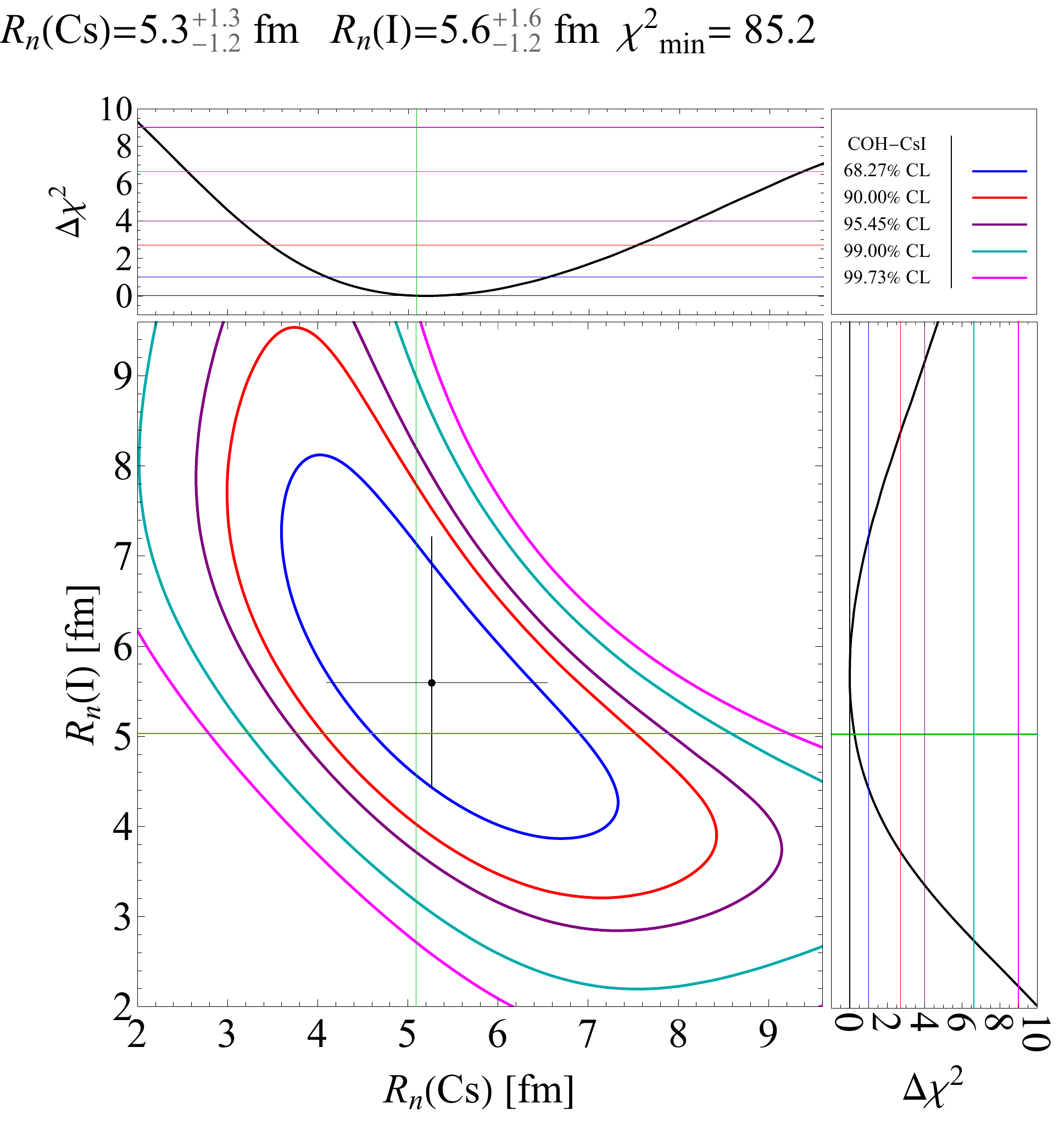}
\\
\end{tabular}
}
\\
\end{tabular}
\caption{ \label{fig:2Dsin2rn}
Constraints obtained fitting the COHERENT CsI data on the weak mixing angle and the average rms CsI neutron radius (a) and on the plane of $R_n(^{133}{\rm{Cs}})$ and $R_n(^{127}{\rm{I}})$ (b)  together with their marginalizations, at different CLs. The green lines indicate the theoretical low-energy value of the weak mixing angle and the NSM prediction for the corresponding rms neutron distribution radius.
}
\end{figure*}

\begin{figure*}[!t]
\centering
\setlength{\tabcolsep}{0pt}
\begin{tabular}{cc}
\subfigure[]{\label{fig:2Dsin2rnAPVa}
\begin{tabular}{c}
\includegraphics*[width=0.42\linewidth]{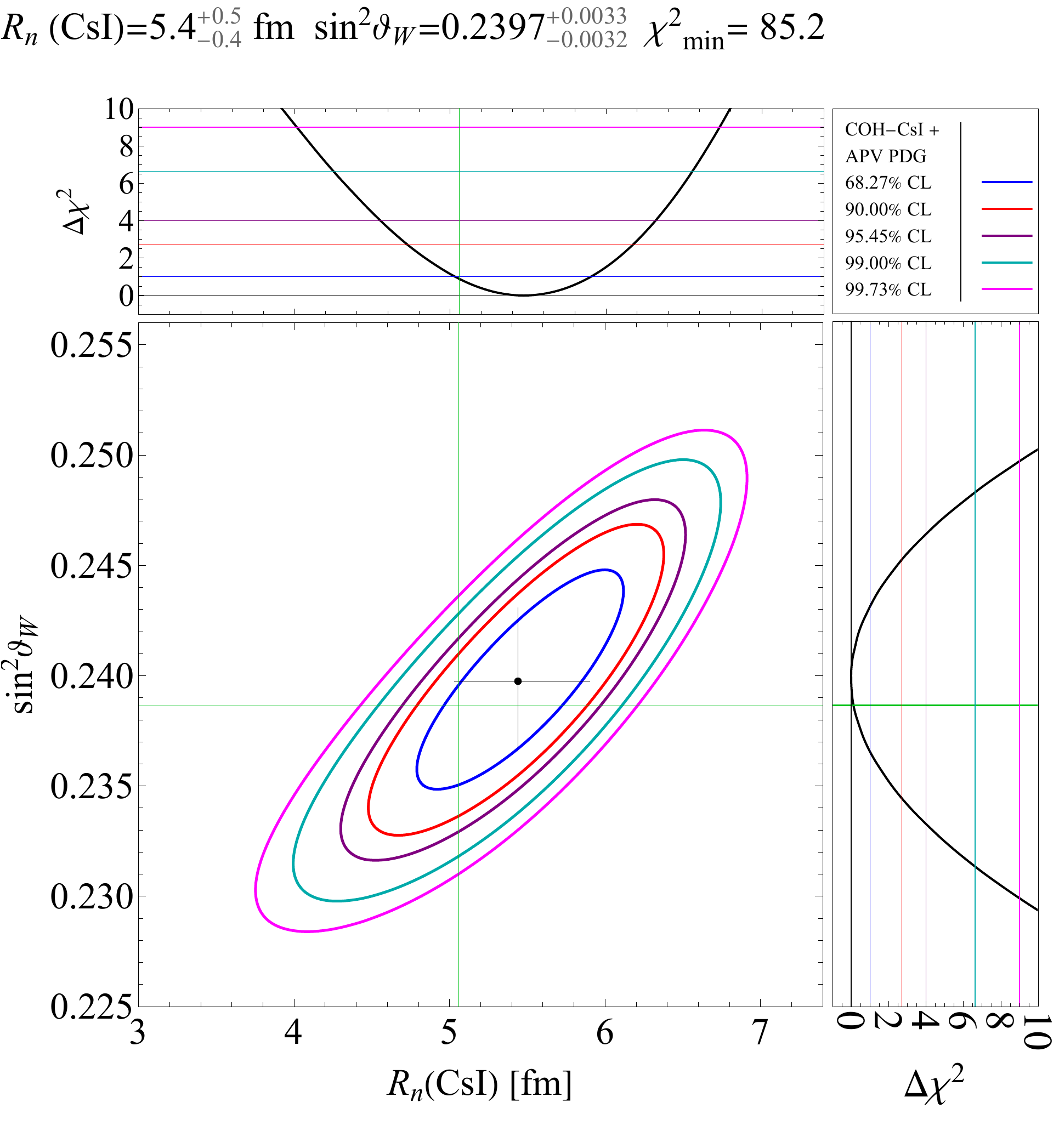}
\\
\end{tabular}
}
&
\subfigure[]{\label{fig:2Dsin2rnAPVb}
\begin{tabular}{c}
\includegraphics*[width=0.42\linewidth]{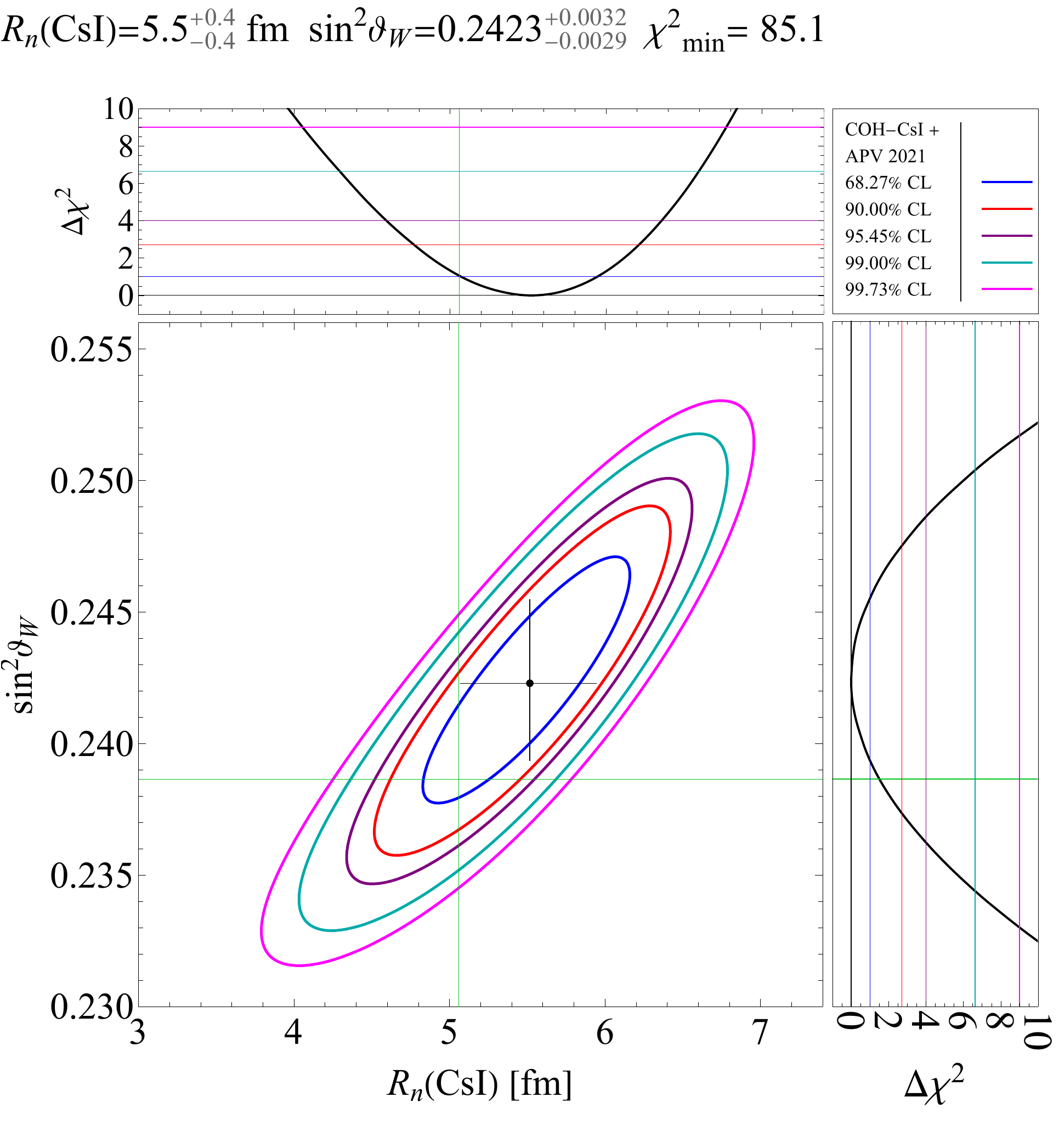}
\\
\end{tabular}
}
\\
\end{tabular}
\caption{ \label{fig:2Dsin2rnAPV}
Constraints on the weak mixing angle and the average rms CsI neutron radius together with their marginalizations, at different CLs obtained fitting the COHERENT CsI data in combination with APV data, using the value for the neutron skin corrections of Ref.~\cite{Dzuba:2012kx} (a) and Ref.~\cite{Sahoo:2021thl} (b). The green lines indicate the theoretical low-energy value of the weak mixing angle and the NSM prediction for the average rms CsI neutron radius.
}
\end{figure*}

In this section, we extend the analysis presented in Sec.~\ref{sec:1dsin2} by letting both the weak mixing angle and the average CsI neutron radius to vary freely in the fit such that correlation between these two observables is properly taken into account. This allows one to obtain simultaneous information on both parameters, taking into account their degeneracy and obtaining thus a more reliable result.
We also perform this fit in combination with APV data using Eq.~(\ref{chicoherentapvRn}), in order to exploit the different dependence of the two experiments on these two parameters. This allows us to get a more precise and solid determination of both quantities that uses the overall constraining power of two different electroweak probes.
The contours at different CLs of the allowed regions in the plane of the weak mixing angle and the average CsI neutron radius are reported in Fig.~\ref{fig:2Dsin2rna}, using COHERENT CsI data alone. At the $1\sigma$ CL we obtain 
\begin{equation}
  {\rm{COH-CsI:}}\;\;\;\sin^2\vartheta_W=0.31^{+0.08}_{-0.07}, \;\; R_n(\mathrm{CsI}) = 6.6^{+1.4}_{-1.1}\;{\rm{fm}}.
\end{equation}
The fit tends to prefer large values for both parameters, with the  theoretical value of the weak mixing angle and the rms average neutron radius of CsI that lie respectively at $\sim1\sigma$ and $\sim1.3\sigma$ outside the marginalized allowed region, despite the large uncertainties.
Indeed, the dataset is better fitted considering unusually large values for the weak mixing angle and the average rms neutron radius. In fact, it is interesting to notice that the fit of the COHERENT-CsI data improves noticeably  with respect to the case in which both parameters are fixed to their theoretical value with a $\Delta\chi^2$ (i.e. the difference between the $\chi^2_{\rm{min}}$ in this fit and the $\chi^2$ obtained fixing the parameters to their theoretical value) which is $\Delta\chi^2=-2.2$.\\

\begin{figure}[h]
    \centering
    \includegraphics[width=0.7\textwidth]{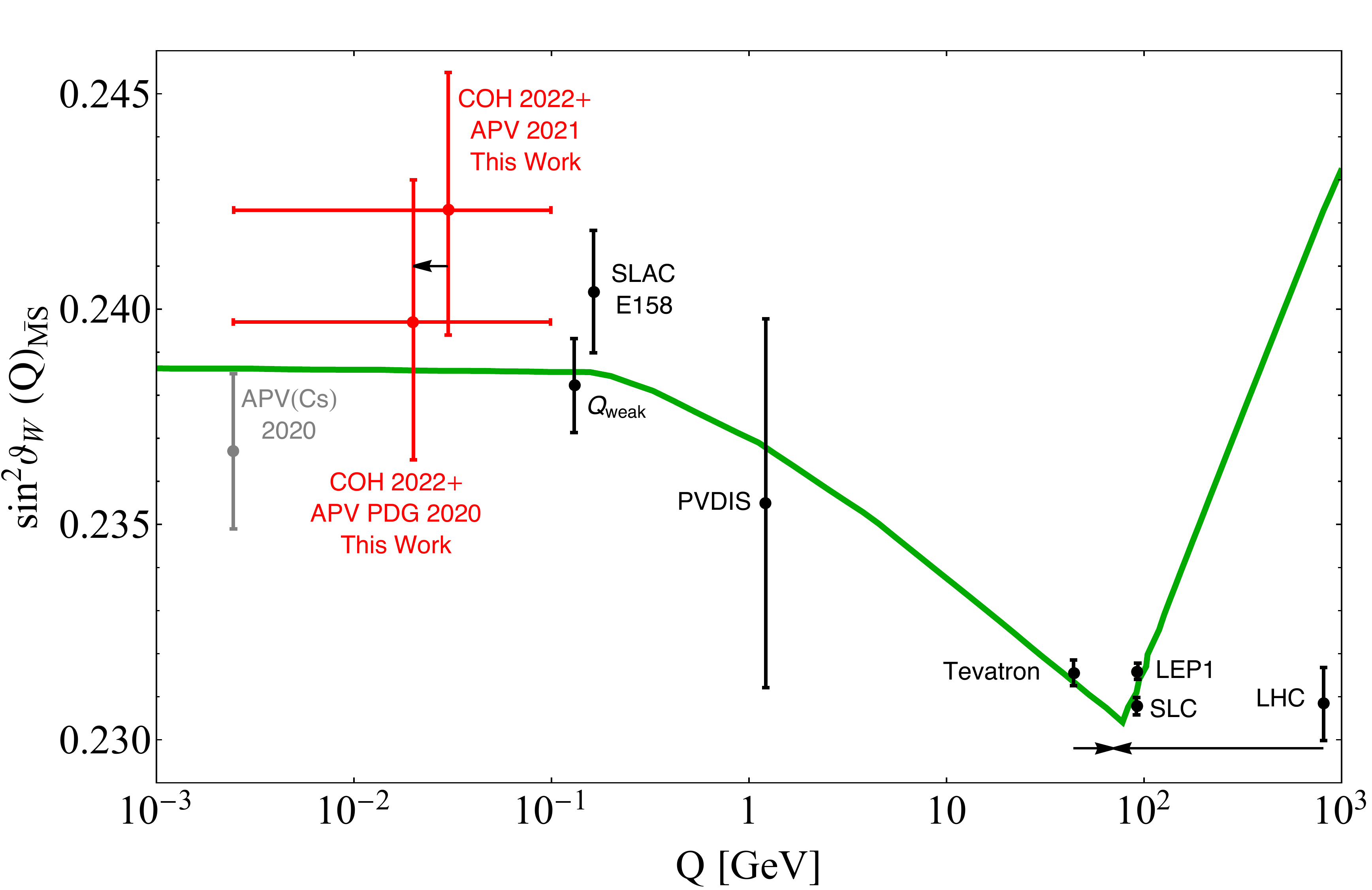}
    \caption{Running of the weak mixing angle in the SM (green line) as a function of the energy scale $Q$. The black experimental determination represent the status of the art of the measurements at different energy scales~\cite{ParticleDataGroup:2022pth,Dzuba:2012kx,Anthony:2005pm,Wang:2014bba,Zeller:2001hh, Androic:2018kni}. The red points show the determinations from the combined analysis of APV(Cs) and COHERENT-CsI measurements retrieved in this work, which supersedes the nominal APV determination depicted in grey~\cite{Wood:1997zq}.}
    \label{fig:Running}
\end{figure}

These constraints have been further tested by performing a combined fit with the APV experiment on cesium.
The results of the combined analyses are reported in Fig.~\ref{fig:2Dsin2rnAPVa} and Fig.~\ref{fig:2Dsin2rnAPVb}, using the experimental value of $Q_W$ obtained with the theoretical prediction of the PNC amplitude of Ref.~\cite{Dzuba:2012kx} and that recently calculated in Ref.~\cite{Sahoo:2021thl}, respectively.
Using $({\rm Im}\, E_{\rm PNC})_{\rm th.}^{\rm w.n.s.}$ from Ref.~\cite{Dzuba:2012kx} we obtain
\begin{equation}
  {\rm{APV\, PDG + COH-CsI:}}\;\;\;\sin^2\vartheta_W=0.2397^{+0.0033}_{-0.0032}, \;\; R_n(\mathrm{CsI}) =5.4^{+0.5}_{-0.4}\;{\rm{fm}}.
\end{equation}
The impact of including the APV data is noticeable, both in the uncertainty of the parameters, that is improved by more that one order of magnitude for the weak mixing angle and by a factor of $\sim3$ for $R_n(\mathrm{CsI})$, as well as in their central values, that are moved towards the expected values, especially for $\textrm{sin}^2\vartheta_W$.\\
Using $({\rm Im}\, E_{\rm PNC})_{\rm th.}^{\rm w.n.s.}$ from Ref.~\cite{Sahoo:2021thl}, at the $1\sigma$ level we obtain
\begin{equation}
  {\rm{APV\, 2021 + COH-CsI:}}\;\;\;\sin^2\vartheta_W=0.2423^{+0.0032}_{-0.0029}, \;\; R_n(\mathrm{CsI}) =5.5^{+0.4}_{-0.4}\;{\rm{fm}}.
\end{equation}
These results are depicted by the red data points in Fig.~\ref{fig:Running}, where a summary of the weak mixing angle measurements as a function of the energy scale Q is shown along with the SM predicted running calculated in the $\overline{\textrm{MS}}$ scheme. They represent an alternative derivation of the weak mixing angle from APV that is fully data-driven and that keeps into account the
correlation with the value of $R_n$ determined simultaneously using two electroweak probes, that are known to
be practically model independent. Indeed, the nominal derivation of the weak mixing angle from APV data, that is reported in the PDG~\cite{ParticleDataGroup:2022pth} and is depicted by the grey point in Fig.~\ref{fig:Running}, uses a value of $R_n$ that is extrapolated from hadronic experiments using antiprotonic atoms, which are known to be affected by considerable model dependencies.
By comparing the two new determinations reported in this work, it is possible to see that  the weak mixing angle is especially affected by the particular choice of the PNC amplitude, underlying thus the importance for the future to clarify the discrepancies between the two different approaches used in Refs.~\cite{Sahoo:2021thl,Dzuba:2012kx}.

\subsection{Simultaneous determination of $\mathbf{R_{n}}$(I) vs $\mathbf{R_{n}}$(Cs) and combined analysis with APV}\label{sec:Csi2DRnIRnCs}

\begin{figure*}[!]
\centering
\setlength{\tabcolsep}{0pt}
\begin{tabular}{cc}
\subfigure[]{\label{fig:2DRnCsRnIa}
\begin{tabular}{c}
\includegraphics*[width=0.42\linewidth]{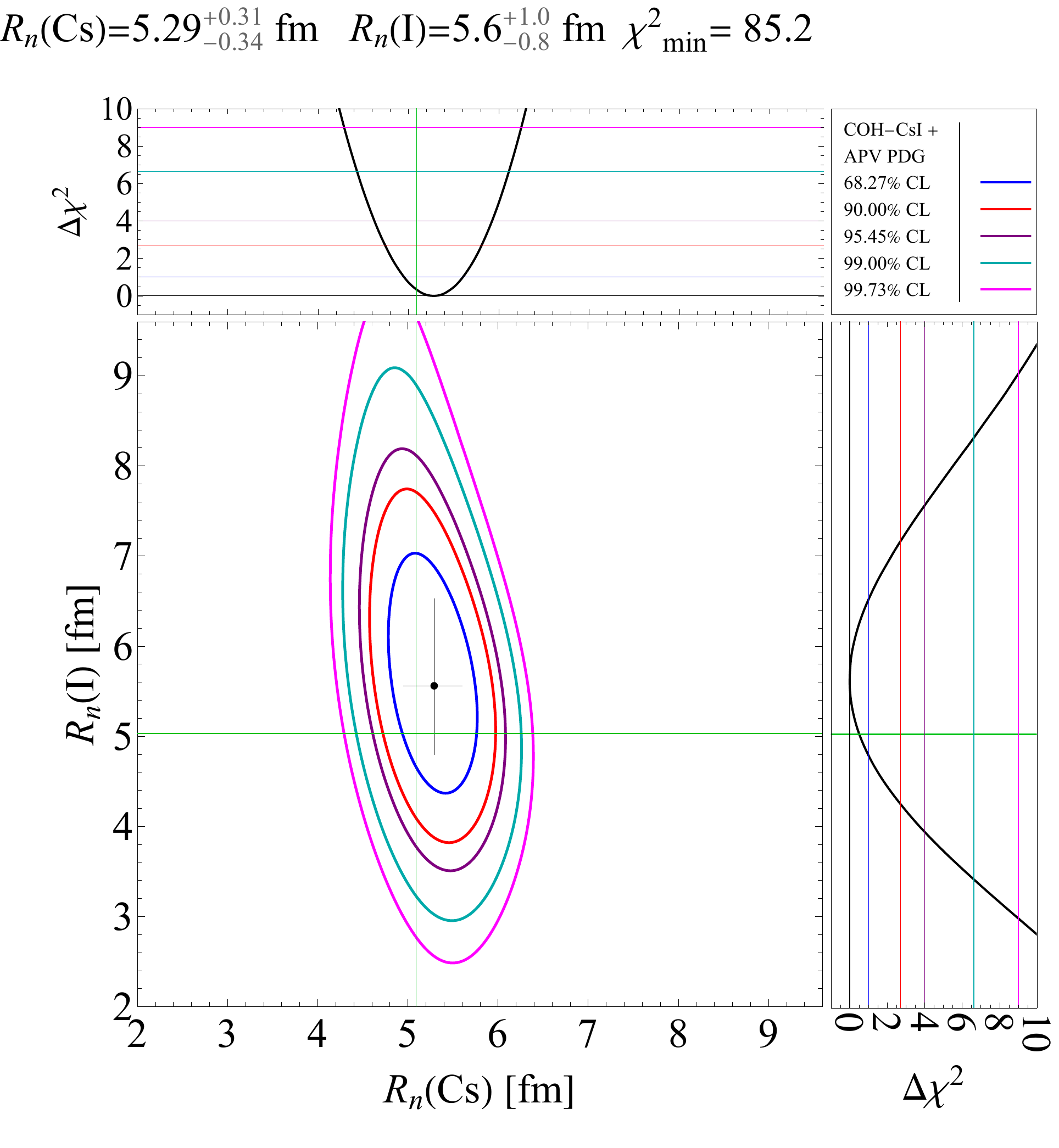}
\\
\end{tabular}
}
&
\subfigure[]{\label{fig:2DRnCsRnIb}
\begin{tabular}{c}
\includegraphics*[width=0.42\linewidth]{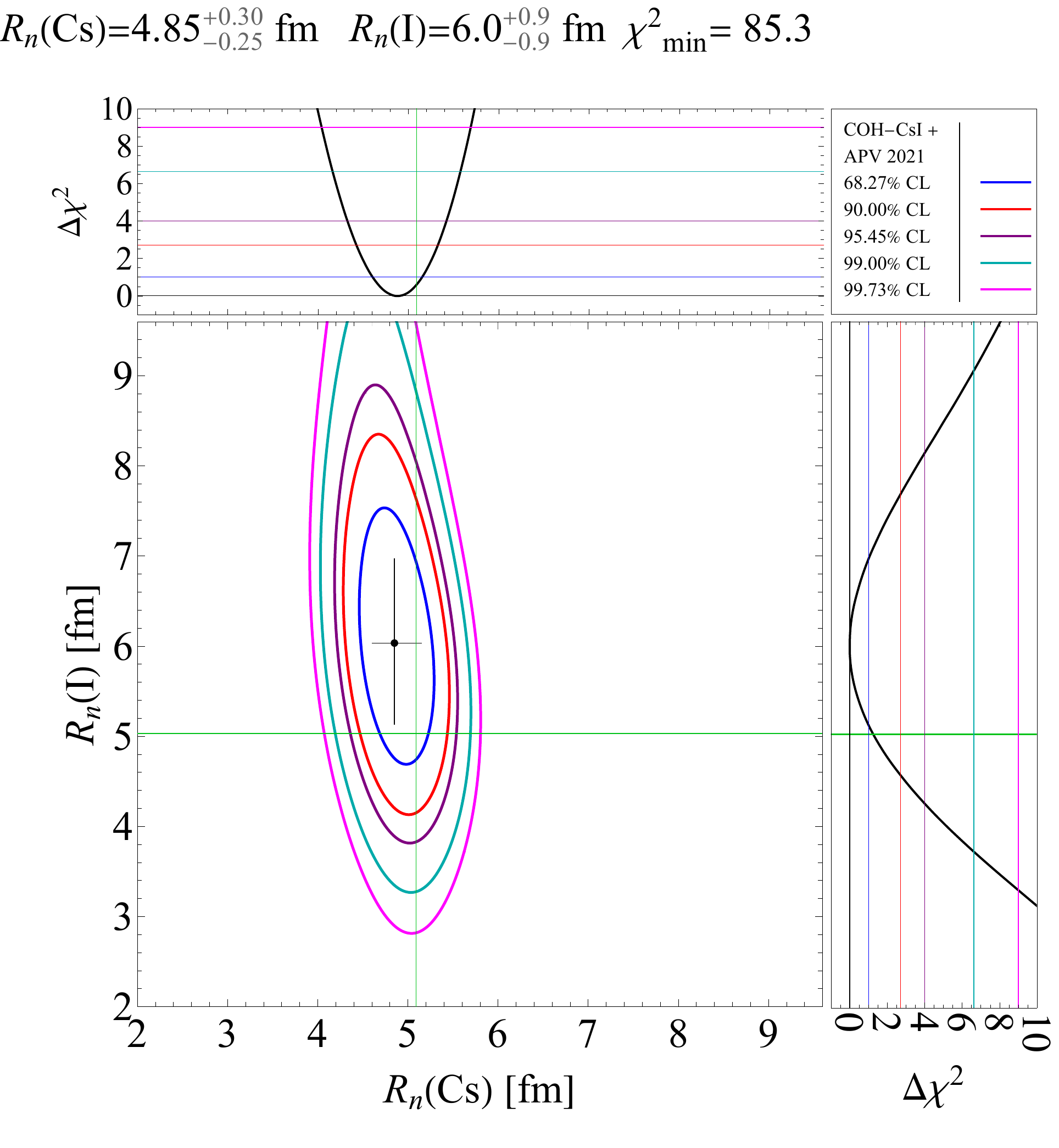}
\\
\end{tabular}
}
\\
\end{tabular}
\caption{ \label{fig:2DRnCsRnI+APV}
Constraints on the plane of $R_n(^{133}{\rm{Cs}})$ and $R_n(^{127}{\rm{I}})$ together with their marginalizations, at different CLs obtained fitting the COHERENT CsI data in combination with APV data, using the value for the neutron skin corrections of Ref.~\cite{Dzuba:2012kx} (a) and Ref.~\cite{Sahoo:2021thl} (b). The green lines indicate the corresponding NSM prediction for the average rms neutron radius of Cs and I.
}
\end{figure*}

In this section, we describe the study of the correlation between $R_n({\rm{Cs}})-R_n({\rm{I}})$ using the latest COHERENT CsI data alone and combined with APV Cs to obtain the up-to-date and more accurate constraints on both these quantities. In fact, since APV depends only on the Cs neutron radius, while the COHERENT CsI result depends on both $R_n{(^{133}{\rm{Cs}})}$ and $R_n{(^{127}{\rm{I}})}$, we are able to break their degeneracy by fixing the weak mixing angle to its SM value.
The result of the COHERENT CsI analysis is reported in Fig.~\ref{fig:2DrnCsrnI}, where we show the contours at different CLs in the plane of $R_n({\rm{Cs}})$ and $R_n ({\rm{I}})$. Namely, we get
\begin{equation}
  {\rm{COH-CsI:}}\;\;\;R_n({\rm{Cs}})=5.3^{+1.3}_{-1.2}\;{\rm{fm}}, \;\; R_n(\mathrm{I}) = 5.6^{+1.6}_{-1.2}\;{\rm{fm}}.
\end{equation}
As expected, COHERENT CsI data alone does not allow to disentangle the two contributions, motivating the need to perform a combined analysis with APV data.
The results of the combined analysis are reported in Fig.~\ref{fig:2DRnCsRnIa} and Fig.~\ref{fig:2DRnCsRnIb}, using the experimental value of $Q_W$ obtained with the theoretical predictions of the PNC amplitude considered in this work.
In these two scenarios we obtain 
\begin{eqnarray}
{\rm{APV\, PDG + COH-CsI:}}\;\;\;R_n(\mathrm{Cs}) &=& 5.29^{+0.31}_{-0.34}\;{\rm{fm}}, \;\; R_n(\mathrm{I}) = 5.6^{+1.0}_{-0.8}\;{\rm{fm}}, \\
{\rm{APV\, 2021 + COH-CsI:}}\;\;\;R_n(\mathrm{Cs}) &=& 4.85^{+0.30}_{-0.25}\;{\rm{fm}}, \;\; R_n(\mathrm{I}) = 6.0^{+0.9}_{-0.9}\;{\rm{fm}}. 
\end{eqnarray}
The corresponding neutron skins are, respectively, $\Delta R_{\rm{np}}({\rm{Cs}})=0.20^{+0.31}_{-0.34}\;{\rm{fm}}$-$\Delta R_{\rm{np}}({\rm{I}})=0.57^{+1.0}_{-0.8}\;{\rm{fm}}$ and $\Delta R_{\rm{np}}({\rm{Cs}})=-0.24^{+0.30}_{-0.25}\;{\rm{fm}}$-$\Delta R_{\rm{np}}({\rm{I}})=1.0^{+0.9}_{-0.9}\;{\rm{fm}}$.
Also in this case the usage of the different PNC amplitudes play a major role and with the second analysis, the slightly larger value of the iodium rms neutron radius is compensated by a significantly smaller value of $R_n(\textrm{Cs})$, that translates in an almost-zero neutron skin for cesium, with smaller uncertainties than those of the first analysis.
Moreover, in all the scenarios, the central values suggest that $R_n(\textrm{I})>R_n(\textrm{Cs})$, while all theoretical models (see e.g. Table I of Ref.~\cite{Cadeddu:2021ijh}) predicts the opposite. We thus redetermine these measurements after imposing the well-motivated constraint $R_n(\textrm{I}) \le R_n(\textrm{Cs})$. In this case the measurements performed in this section become
\begin{eqnarray}
{\rm{COH-CsI}}\, [R_n(\textrm{I}) \le R_n(\textrm{Cs})] &:&\;\;\;R_n({\rm{Cs}})=5.5^{+1.1}_{-0.4}\;{\rm{fm}}, \;\; R_n(\mathrm{I}) = 5.4^{+0.4}_{-1.0}\;{\rm{fm}},\\
{\rm{APV\, PDG + COH-CsI}}\, [R_n(\textrm{I}) \le R_n(\textrm{Cs})] &:&\;\;\;R_n(\mathrm{Cs}) = 5.32^{+0.30}_{-0.23}\;{\rm{fm}}, \;\; R_n(\mathrm{I}) = 5.30^{+0.30}_{-0.6}\;{\rm{fm}}, \\
{\rm{APV\, 2021 + COH-CsI}}\, [R_n(\textrm{I}) \le R_n(\textrm{Cs})] &:&\;\;\;R_n(\mathrm{Cs}) = 5.07^{+0.21}_{-0.26}\;{\rm{fm}}, \;\; R_n(\mathrm{I}) = 5.06^{+0.22}_{-0.4}\;{\rm{fm}}. 
\end{eqnarray}
Imposing this constraint, we achieve an uncertainty as low as 4\% on $R_n(\textrm{Cs})$.
The corresponding constraints on the plane of $R_n(^{133}{\rm{Cs}})$ and $R_n(^{127}{\rm{I}})$ together with their marginalizations, at different CLs can be found in Appendix~\ref{app:RnIRnCs}.

\subsection{Overview of the results on $\mathbf{R_n}$}\label{sec:overview}

\begin{figure}[h]
    \centering
\centering
\setlength{\tabcolsep}{0pt}
\begin{tabular}{cc}
\subfigure[]{\label{fig:overviewPDG}
\begin{tabular}{c}
\includegraphics*[width=0.8\linewidth]{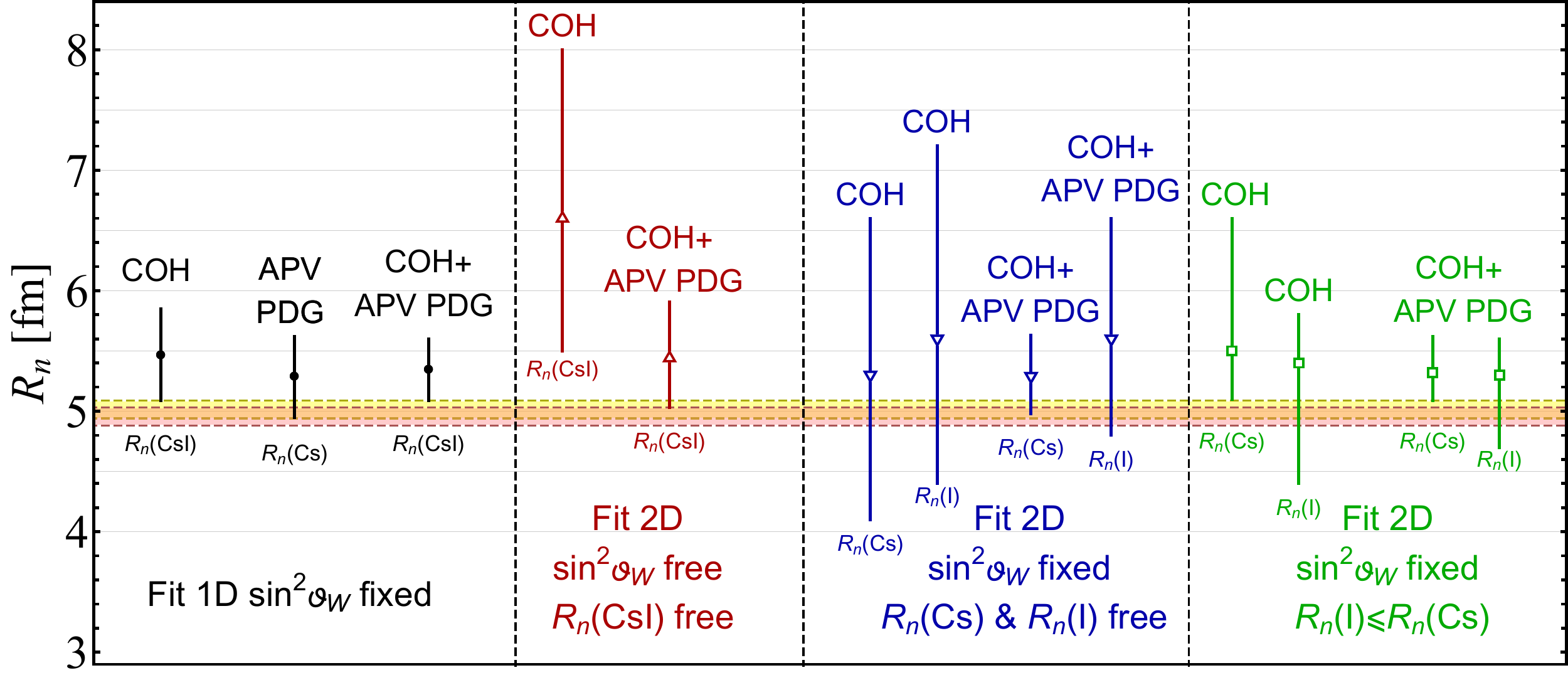}
\\
\end{tabular}
}
\end{tabular}
\begin{tabular}{cc}
\subfigure[]{\label{fig:overview2021}
\begin{tabular}{c}
\includegraphics*[width=0.8\linewidth]{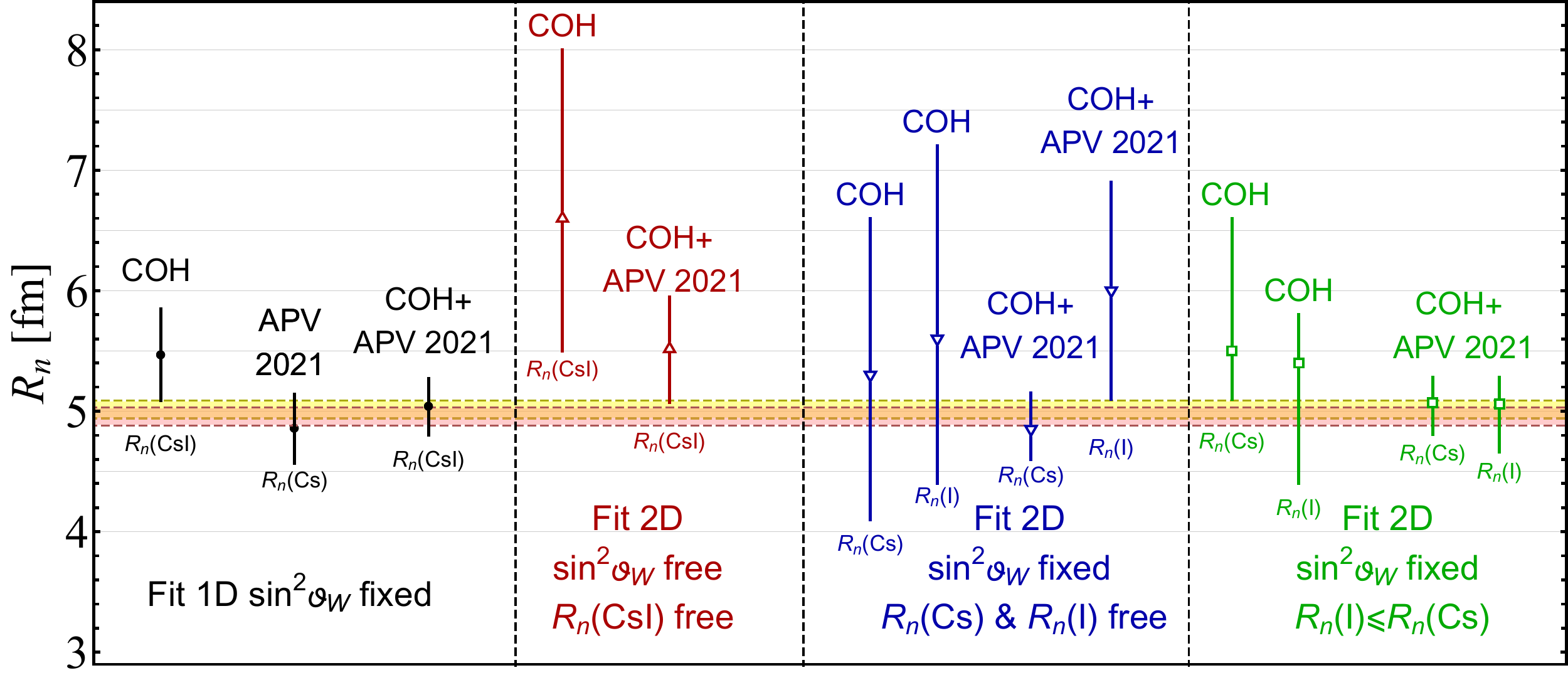}
\\
\end{tabular}
}
\\
\end{tabular}
    \caption{Overview of the different results presented in this work obtained using \cenns CsI COHERENT (COH) data and APV with the PNC amplitude (a) from Ref.~\cite{Dzuba:2012kx} (APV PDG) or (b) from Ref.~\cite{Sahoo:2021thl} (APV 2021) as well as their combination on $R_n(\mathrm{CsI})$ and $R_n(\mathrm{Cs})$ when fixing the weak mixing angle to $\hat{s}^2_0$ (black circles), on $R_n(\mathrm{CsI})$ when fitting simultaneously for $R_n$ and $\sin^2 \vartheta_{\text{W}}$ (red up triangles), on $R_n(\mathrm{Cs})$ and $R_n(\mathrm{I})$ when fixing the weak mixing angle to $\hat{s}^2_0$ (blue down triangles) as well as when imposing the constraint $R_n(\mathrm{Cs})\geq R_n(\mathrm{I})$ (green squares). The yellow and orange areas represent the regions where the theoretical predictions of $R_n(\mathrm{Cs})$ and $R_n(\mathrm{I})$ can be found, as taken from Table I of Ref.~\cite{Cadeddu:2021ijh} and from NSM calculations in Ref.~\cite{Hoferichter:2020osn}.}
    \label{fig:overview}
\end{figure}

Given the vast amount of measurements of the neutron rms radius distribution presented in this work under different hypotheses, we summarised all of them in Fig.~\ref{fig:overview}(a) and (b) when using APV with the PNC amplitude from Ref.~\cite{Dzuba:2012kx} or from Ref.~\cite{Sahoo:2021thl}, respectively. Despite the different fit configurations used to extract the values of $R_n(\mathrm{CsI})$, $R_n(\mathrm{Cs})$ and $R_n(\mathrm{I})$, a coherent picture emerges with an overall agreement between the COHERENT and APV results and the theoretical predictions. However, we would like to note that using APV PDG we obtain on average larger values on the radii, even if still compatible within uncertainties. On the contrary, APV 2021 shifts downwards the measured radii towards the predictions, but in the simultaneous 2D fit with $\sin^2 \vartheta_{\text{W}}$ where the correlation with the latter increases the extracted central value of $R_n(\mathrm{CsI})$. Moreover, we checked the impact of using a different quenching factor, by comparing our nominal results obtained using Refs.~\cite{COHERENT:2021xmm,COHERENT:2021pcd} and the derivation in Ref.~\cite{Lewis:2021cjv}. The latter lower QF decreases the total number of \cenns events resulting in a smaller $R_n(\mathrm{CsI})$ by about 10\%.

\section{Future perspectives}
\label{sec:future}

\begin{figure}[h]
    \centering
    \includegraphics[width=\textwidth]{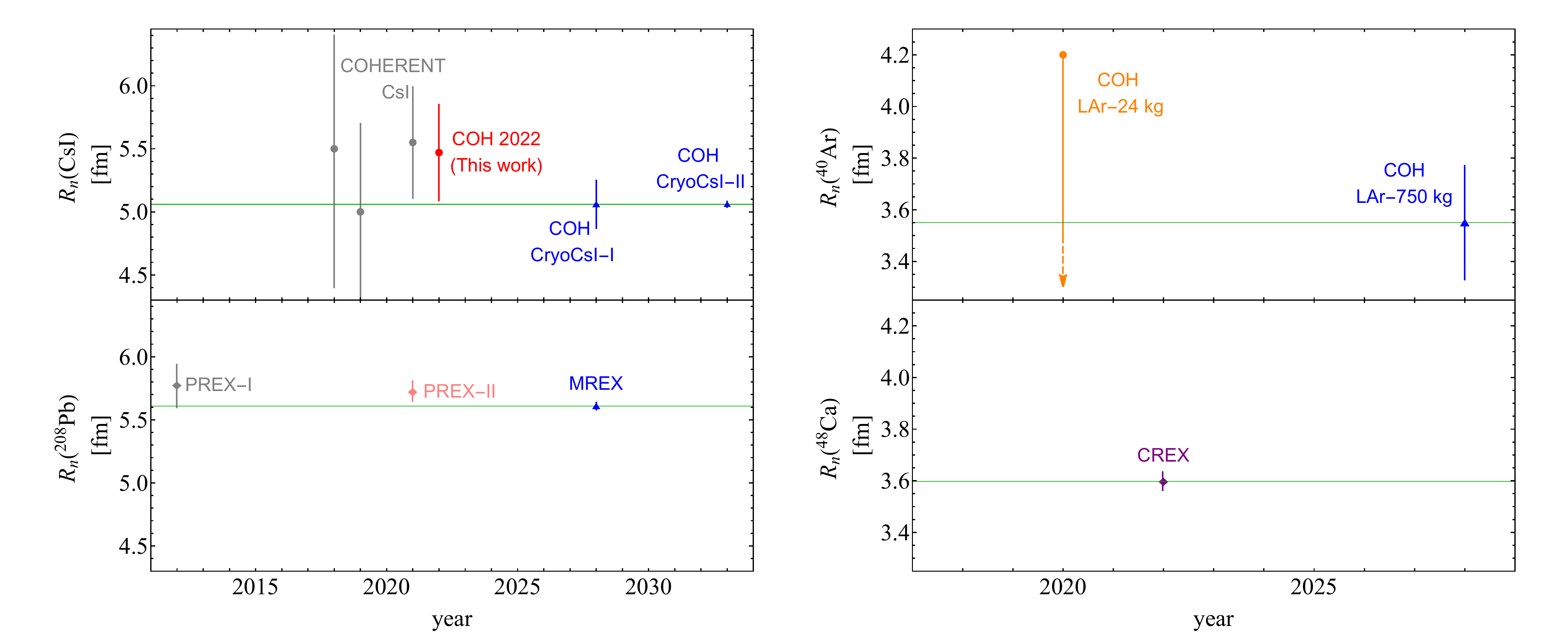}
    \caption{Current status and future projections for neutron distribution radii of different nuclei measured via electroweak probes. The upper plots show the current and foreseen measurements of $R_n(\mathrm{Cs}$ from \cenns with CsI crystal detectors~\cite{Cadeddu:2017etk,Cadeddu:2019eta,Cadeddu:2021ijh} (left) and of $R_n(\mathrm{Ar}$ liquid Ar detectors~\cite{Cadeddu:2020lky,Akimov:2022oyb} (right), compared to the lower plots where the current and foreseen measurements from parity-violating electron scattering are shown, for the case of Pb~\cite{Abrahamyan:2012gp,PREX:2021umo} (left) and Ca~\cite{CREX:2022kgg} (right).  For CsI, similar uncertainties are expected to be achieved thanks to the detectors planned at the ESS~\cite{Abele:2022iml} and at the CSNS~\cite{Su:2023klh}.} See the text for more details.
    \label{fig:SummaryPlot}
\end{figure}

In this section, we describe a sensitivity study that we performed to outline the potentialities of \cenns to measure the CsI neutron radius and the weak mixing angle at the SNS, in the context of the COHERENT experimental program.
As shown in the previous sections and in several previous papers
(see, e.g., the review in Ref.~\cite{Abdullah:2022zue}),
\cenns is a very powerful tool,
being a process which turned out to be very versatile in putting constraints on a variety of parameters.
Nevertheless, the current level of accuracy of \cenns in the determination of the neutron radius, both the CsI one reported in this work as well as the argon (Ar) one reported in Ref.~\cite{Cadeddu:2020lky}, is still lower with respect to that obtained using parity-violating electron scattering on similar nuclei. This is visible in Fig.~\ref{fig:SummaryPlot} where we show the current status for different neutron distribution radii measured via diverse electroweak probes. As shown, the precision achieved by the PREX~\cite{Abrahamyan:2012gp,PREX:2021umo,Becker:2018ggl} and CREX~\cite{CREX:2022kgg} experiments is indeed greater than that obtained through \cenns for CsI and Ar~\cite{Cadeddu:2017etk,Cadeddu:2019eta,Cadeddu:2021ijh,Cadeddu:2020lky,Akimov:2022oyb}. 

Luckily, this is not the end of the story. The COHERENT collaboration has additional existing and planned near-future deployments in the Neutrino Alley at the SNS with exciting physics potential. In particular, the experimental program under development includes a tonne-scale liquid argon time-projection chamber detector as well as a large scale CsI cryogenic detector. These new detectors, together with planned upgrades to the SNS proton beam, will further broaden and deepen the physics reach of the COHERENT experiment.  Moreover, the European Spallantion Source (ESS) is currently under construction in Lund, Sweden~\cite{Abele:2022iml}. At design specifications, the ESS will operate at 5 MW using a proton linac with a beam energy of 2 GeV. In
addition to providing the most intense neutron beams, the ESS also provides a large neutrino flux and so it will be exploited to study \cenns, using in particular a 31.5 kg CsI target kept at 80~K~\cite{Abele:2022iml,Baxter:2019mcx}. Finally, a new \cenns detection experiment is under construction in China, where undoped CsI crystals coupled with two photon multiplier tubes each, will be cooled down to 77~K and placed at the China Spallation Neutron Source (CSNS)~\cite{Su:2023klh}.\\

In this work, we performed a sensitivity study using COHERENT plans as described in Ref.~\cite{Akimov:2022oyb}. However, similar conclusions and prospects can also be drawn for the already mentioned CsI detectors expected at the ESS and the CSNS as they foresee similar technologies.  The aim of this study is to assess to which extent \cenns will be competitive in the future. 
To better appreciate the reach of the \cenns program, we compare our COHERENT projections with other competitive measurements that are foreseen for the neutron radius of different nuclei and for the weak mixing angle. In this way, we obtain a complete and comprehensive review of the current and future status of knowledge on these fundamental quantities.
The first upgrade of the SNS is planned for 2025, where the proton beam energy $(E_p)$ will be increased up to $1.3\;\rm{GeV}$ with respect to the current $0.984\;\rm{GeV}$. Moreover, the beam power $P_{\rm{beam}}$ will increase to $2\;\rm{MW}$, compared to the current $1.4\;\rm{MW}$ so that the number of neutrinos per flavor  produced for each proton-on-target (POT) will increase to a value of 0.12. A second target station is planned in the 2030s, for a final power of $2.8\;\rm{MW}$.
Using this information, we evaluate the number of proton-on-target ($N_{\rm{POT}}$) expected, that describe the intensity of the neutrino flux (see e.g. Ref.~\cite{AtzoriCorona:2022moj}), which is given by
\begin{equation}
   N_{\rm{POT}}=t_{\rm{exp}}\frac{P_{\rm{beam}}}{E_p},
\end{equation}
$t_{\rm{exp}}$ being the running time of the experiment.
This means that the future SNS upgrade will be able to get a much higher neutrino flux for each neutrino flavor with respect to the current configuration. \\

The so-called COH-CryoCsI-I experiment, scheduled for 2025, will have a mass of about 10 kg and will exploit an undoped CsI crystal at cryogenic temperature $(\sim40\rm{K})$, which would permit to use SiPM arrays instead of PMTs in order to remove the Cherenkov radiation background emitted by the latter. Moreover, the undoped CsI crystals at cryogenic temperature have approximately twice the light yield of the CsI[Na] crystal at $300~\rm{K}$.
The following upgrade will be the COH-CryoCsI-II experiment, planned in the 2030s, that will operate in similar conditions with a 700~kg undoped CsI detector. Both the COH-CryoCsI-I and COH-CryoCsI-II detectors will be able to lower the energy threshold, which is a fundamental requirement for \cenns precision physics. Following Ref.~\cite{Akimov:2022oyb}, we considered a threshold of $1.4\;\rm{keV}_{nr}$ while keeping the shape of the energy efficiency unchanged.
In addition, the systematic uncertainty on the neutrino flux will be strongly reduced thanks to the planned $\rm{D}_2 O$ detector and will approach 4.7(2)$\%$ statistical uncertainty after 2(5) SNS-years of operation. In this sensitivity study, we will use the value of 4.7$\%$ for the statistical uncertainty on the neutrino flux for the Cryo-CsI-I detector, while we use 2$\%$ for the Cryo-CsI-II detector. Another fundamental ingredient is the quenching factor of the CsI crystal, whose behavior at cryogenic temperature is currently under investigation. It will be directly measured by the COHERENT collaboration, but for this sensitivity study, we will consider a quenching factor of 5$\%$ as reported in Ref.~\cite{Akimov:2022oyb} and as it has already been established in Ref.~\cite{Lewis:2021cjv}. Given the conservative value used, we do not consider a systematic error on the quenching factor. Thus, the total systematic uncertainty considered for the \cenns prediction is $\sigma_{\rm{CE}\nu\rm{NS}}=0.062$ for the Cryo-CsI-I detector and $\sigma_{\rm{CE}\nu\rm{NS}}=0.046$ for the Cryo-CsI-II detector.
Assuming three years of data taking, $\sim2\cdot10^3$ and $\sim2\cdot10^5$ events are expected for COH-CryoCsI-I and COH-CryoCsI-II\footnote{For comparison, at the ESS a compact 31.5 kg cryogenic CsI detector is expected to measure $\sim1.2\cdot10^4$ \cenns events per year~\cite{Abele:2022iml}.}, respectively, to be compared with the current CsI available statistics of $\sim 300$ events. Given the high statistics of these detectors, we find that the choice of using a poissonian or a gaussian definition for the least square function leads to almost identical results. Moreover, in order to keep into account the timing information in the sensitivity study, we used the least-square function definition in Eq.~(\ref{chi2coherentCsI}), using the same time binning of the data and time efficiency of the latest CsI data release analyzed in this paper. In this sensitivity study, we considered the SS background obtained by rescaling the one measured by the current CsI detector for the exposure time and the mass of COH-CryoCsI-I and COH-CryoCsI-II, respectively. Given that the SS background is unknown for energies smaller than the actual COHERENT energy threshold, we simply extended the background currently measured by the CsI COHERENT detector also to the first energy bin.
\\

In Fig.~\ref{fig:SummaryPlot}, we report the projections on the neutron radius of CsI as obtained from this sensitivity study, along with a comprehensive review of the current and future status of different neutron radius measurements on various nuclei using other weak probes.
We find that COH-CryoCsI-I will be able to measure the neutron rms CsI radius with a precision of $\sigma(R_n\textrm{(CsI)})=0.19\,\rm{fm}$ corresponding to a relative accuracy of about $4\%$\footnote{For these studies, we consider a value of  $R_n\textrm{(CsI)}=5.06\;\rm{fm}$ as provided by the NSM calculations. See Sec.~\ref{sec:theo} for more information.}.
Similarly, for the COH-CryoCsI-II scenario we obtain a sensitivity projection for $R_n\rm{(CsI)}$  that corresponds to $\sigma(R_n\textrm{(CsI)})=0.023\,\rm{fm}$, meaning that COH-CryoCsI-II will be able to reach a per-mille accuracy level, i.e., about $0.5\%$. It is worth noticing that, in this regime, the projected uncertainty on the neutron radius will become smaller than the difference between the Cs and I radius, expected to be $\sim0.06\,\rm{fm}$. Thus, it will be of paramount importance to keep into account the different contributions of Cs and I by performing a simultaneous fit on these two quantities as we did in Sec.~\ref{sec:Csi2DRnIRnCs}. 
This precision, which is also expected to be achieved at the ESS with a similar amount of foreseen \cenns events, will represent an unprecedented window into nuclear physics, making \cenns very competitive with respect to the other weak probes. Specifically, the other available and currently world-leading measurements on the neutron radius of heavy and neutron-rich nuclei come from parity-violating electron scattering as shown in the lower left panel of Fig.~\ref{fig:SummaryPlot}, for the case of $^{208}\rm{Pb}$~\cite{Abrahamyan:2012gp,PREX:2021umo} as measured by PREX-I and PREX-II, respectively. It is worth noticing that the MREX experiment~\cite{Becker:2018ggl} also plans to measure the $^{208}\rm{Pb}$ neutron radius with an accuracy of about $0.5\%$. 
Talking about lighter nuclei, the currently available measurements come from COHERENT for $^{40}\rm{Ar}$~\cite{COHERENT:2020iec} and for $^{48}\rm{Ca}$~\cite{CREX:2022kgg} from the CREX experiment exploiting again parity-violating electron scattering. The COHERENT collaboration foresees to measure $R_n(\rm{Ar})$ to 4.6$\%$ with the upgraded tonne scale argon detector (see also Ref.~\cite{Sierra:2023pnf}), so-called COH-Ar-750~\cite{Akimov:2022oyb}, as shown in the upper right panel of Fig.~\ref{fig:SummaryPlot}. \\

\begin{figure}[h]
    \centering
    \includegraphics[width=\textwidth]{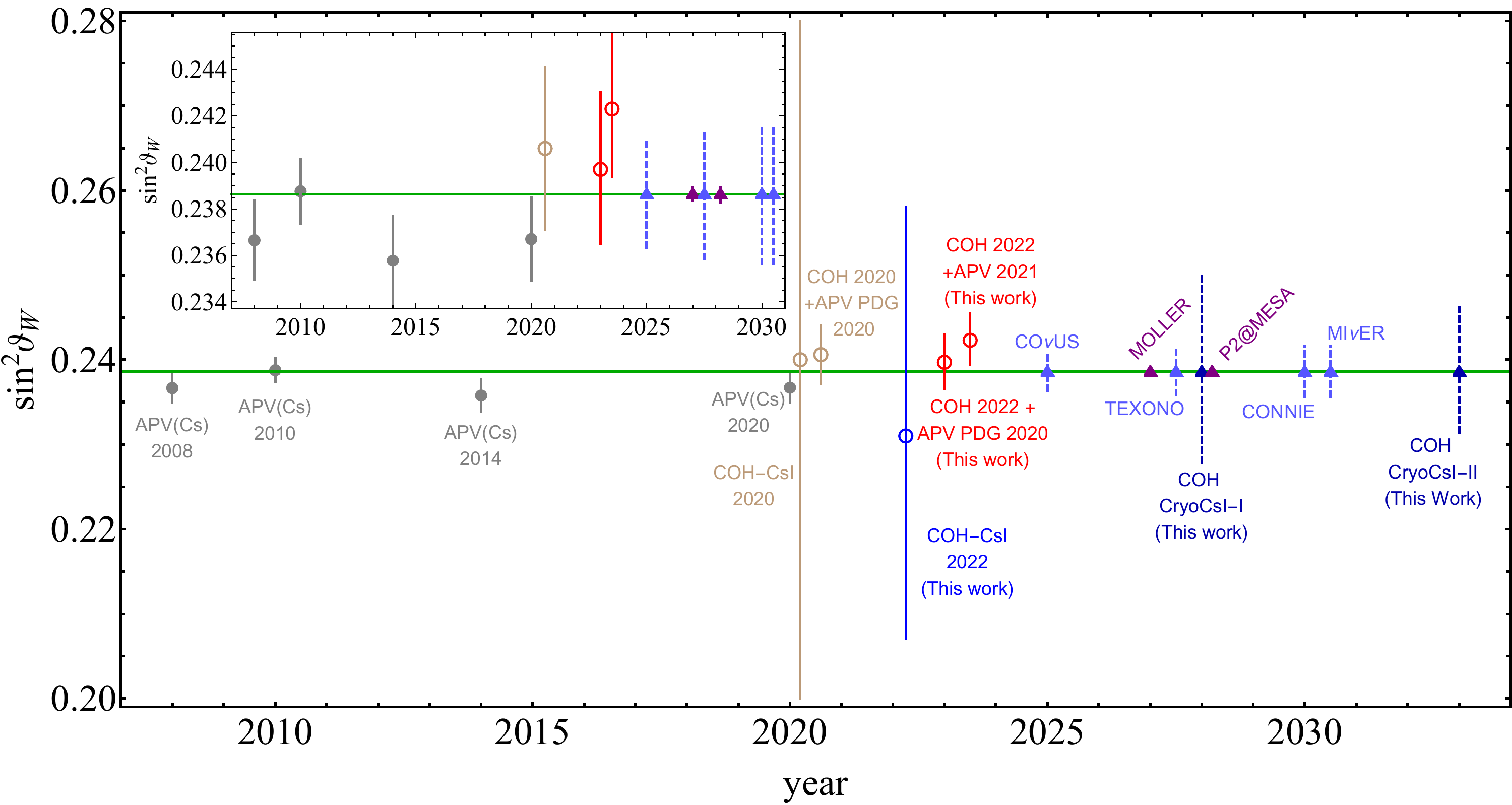}
    \caption{Current status and future projections for weak mixing angle measurements below $Q\lesssim 100\ \mathrm{MeV}$. The gray points show the measurements from APV on cesium atoms during the years~\cite{ParticleDataGroup:2008zun,ParticleDataGroup:2010dbb,ParticleDataGroup:2014cgo,ParticleDataGroup:2020ssz,ParticleDataGroup:2022pth}. The brown measurements  refer to the COHERENT only and the combination between COHERENT and APV as determined in 2020 in our previous work (see Ref.~\cite{Cadeddu:2021ijh}), while the dark blue and red points refer to the updated measurements reported in this work. The projections for future \cenns experiments (CO$\nu$US~\cite{Lindner:2016wff}, TEXONO~\cite{TEXONO:2006xds}, CONNIE~\cite{CONNIE:2016nav}, and MI$\nu$ER~\cite{MINER:2016igy}), shown by the light blue triangles and dashed error bars, are taken from Ref.~\cite{Canas:2018rng}. The purple triangles are the projections for the future electron scattering experiment MOLLER~\cite{Benesch:2014bas} and P2@MESA~\cite{Becker:2018ggl,Dev:2021otb}. The dark blue triangles with dashed error bars are the projections for the future CryoCsI-I and CryoCsI-II determinations as derived in this work. Similar uncertainties are expected to be achieved thanks to the detectors planned at the ESS~\cite{Abele:2022iml} and at the CSNS~\cite{Su:2023klh}.} In the inset in the top left, a zoom of the y axes is shown to better appreciate the statistical uncertainties of the reported measurements and projections, removing the measurements from COHERENT-only which suffer from larger uncertainties.
    \label{fig:SummaryPlots2w}
\end{figure}

In Fig.~\ref{fig:SummaryPlots2w}, we report the projections on the weak mixing angle as obtained from this sensitivity study, along with a comprehensive review of the current and future measurements that are known for an energy scale below $100\ \mathrm{MeV}$. In particular we depicted the evolution of the APV determination in the last decade~\cite{ParticleDataGroup:2008zun,ParticleDataGroup:2010dbb,ParticleDataGroup:2014cgo,ParticleDataGroup:2020ssz}, that moved significantly due to different theoretical re-evaluations, the value of the weak mixing angle extracted from the CsI COHERENT in Ref.~\cite{Cadeddu:2021ijh} and in this work as well as the values obtained from the combination with APV. As it can be seen, many measurements of $\sin^2 \vartheta_{\text{W}}$ are expected in the near future in the low energy sector, as those coming from the P2~\cite{Becker:2018ggl,Dev:2021otb} and MOLLER~\cite{Benesch:2014bas} experiments, and from future \cenns experiments (CO$\nu$US~\cite{Lindner:2016wff}, TEXONO~\cite{TEXONO:2006xds}, CONNIE~\cite{CONNIE:2016nav}, and MI$\nu$ER~\cite{MINER:2016igy}) that will be really powerful for further constraining such a quantity. It is worth noticing that \cenns from reactor antineutrinos already proved to be able to provide a determination of the weak mixing angle. Indeed, this has been shown in Ref.~\cite{AtzoriCorona:2022qrf,AristizabalSierra:2022axl}, even if the uncertainty so far is still too large to be depicted in Fig.~\ref{fig:SummaryPlots2w}. In this scenario, \cenns determinations with CsI will help with both the Cryo-CsI-I detector that will reach a precision of about $\sigma(\sin^2\vartheta_W)=0.012$ and in particular with COH-CryoCsI-II, where a precision of about $\sigma(\sin^2\vartheta_W)=0.007$ will be achieved. Similar precision are also expected to be achieved by the other large cryogenic CsI targets measuring \cenns as highlighted in this section. The sensitivity projection on the weak mixing angle for the Cryo-CsI-I detector has been reported also in Ref.~\cite{Akimov:2022oyb} by the COHERENT collaboration, where a slightly better precision corresponding to $\sigma(\sin^2\vartheta_W)\sim0.009$ has been found.  The different result can be explained considering that the sensitivity to the weak mixing angle depends strongly on the values of $R_n(\rm{Cs})$ and $R_n(\rm{I})$ used to describe the loss of coherence for increasing recoil energies. The values from the NSM calculations adopted in our work differ from the significantly larger value used in the aforementioned work, which seems to be $R_n(\mathrm{CsI})\sim6\;$fm. We verified that we are able to obtain a better agreement with their projections using the latter value for the nuclear radius. When more data will become available, it will be therefore essential to perform a simultaneous determination of these parameters, as investigated in this work.

\section{Conclusions}\label{sec:conclusions}

Motivated by the recent update of the observation of \cenns using a 14.6 kg CsI detector by the COHERENT collaboration, we provide in this manuscript a complete and in-depth legacy determination of the average neutron rms radius of $^{133}\mathrm{Cs}$ and $^{127}\mathrm{I}$ and of the weak mixing angle. To do so, we take advantage of the knowledge developed in the latest years, employing the most up-to-date and accurate description of the \cenns data, profiting from a precise determination of the radiative corrections, the inclusion of the neutrino arrival time information, a refined quenching factor derivation, and an appropriate least-square function definition. Interestingly, we show that the COHERENT CsI data show a 6$\sigma$ evidence of the nuclear structure suppression of the full coherence. Moreover,  we also perform the combination with the APV experimental result, that allows us to disentangle the contributions of the $^{133}\mathrm{Cs}$ and $^{127}\mathrm{I}$ nuclei. A precision as low as $\sim$5(4)$\%$ is obtained on $R_n(\mathrm{Cs})$ leaving it free to vary in the fit (imposing the constraint $R_n(\mathrm{Cs})\geq R_n(\mathrm{I})$).
The combination of APV+COHERENT impacts in particular the determination of the weak mixing angle and allows us to obtain a data-driven measurement of the low-energy weak mixing angle with a percent uncertainty, independent of the value of the average neutron rms radius of $^{133}\mathrm{Cs}$ and $^{127}\mathrm{I}$, that is allowed to vary freely in the fit. 
In all the APV-related results, we exploit two different derivations of the theoretical PNC amplitude, showing that the particular choice can make a difference as large as 6.5\% on $R_n(\mathrm{Cs})$ and 11\% on the weak mixing angle, underlying thus the importance for the future to clarify the discrepancies between the two different approaches used in Refs.~\cite{Sahoo:2021thl} and~\cite{Dzuba:2012kx}. 
At the time of writing, the CsI COHERENT detector has been dismantled, but a new data taking with an upgraded experiment is foreseen in 5 years.
Thus, in light of the recent announcement of a future deployment of a 10~kg and a $\sim$700~kg cryogenic CsI detectors, we provide future prospects for these measurements thanks to a sensitive study that we performed considering all the foreseen improvements. We compared the uncertainties forecasted for $R_n(\mathrm{CsI})$ and $\sin^2\vartheta_W$, which are $\sigma(R_n\textrm{(CsI)})=0.023\,\rm{fm}$ and $\sigma(\sin^2\vartheta_W)=0.007$ for the $\sim$700~kg configuration, with different future measurements of the same quantities using weak probes, highlighting thus the impact that future \cenns measurements will provide.

\appendix
\section{Contours and marginalizations of $R_n(^{133}{\rm{Cs}})$ and $R_n(^{127}{\rm{I}})$ with the constraint $R_n({\rm{Cs}})\geq R_n({\rm{I}})$}
\label{app:RnIRnCs}

\begin{figure*}[ht]
\centering
\setlength{\tabcolsep}{0pt}
\begin{tabular}{cc}
\subfigure[]{\label{fig:RnCsRnICons}
\begin{tabular}{c}
\includegraphics*[width=0.4\linewidth]{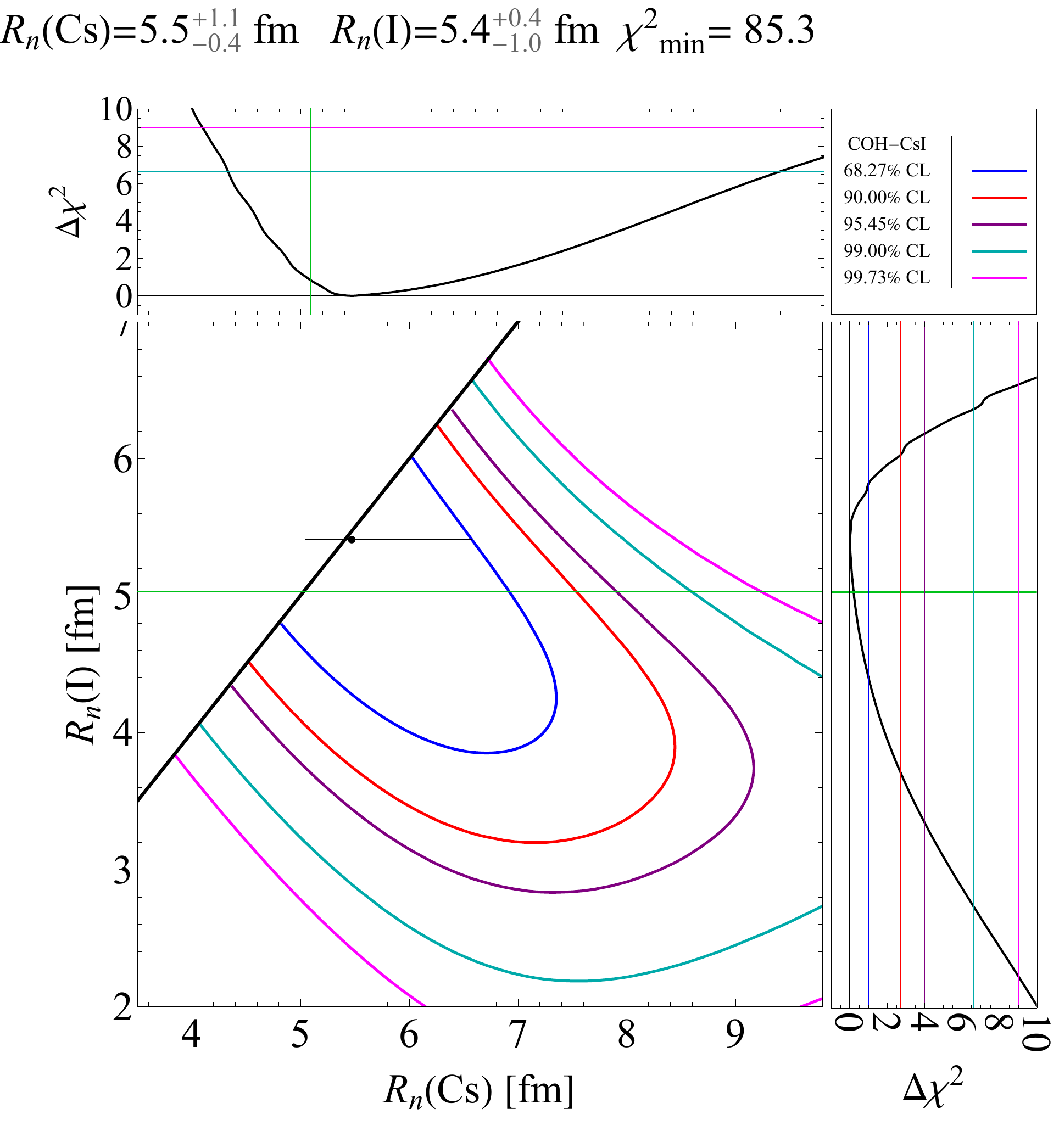}
\\
\end{tabular}
}
&
\subfigure[]{\label{fig:2DRnCsRnIaCons}
\begin{tabular}{c}
\includegraphics*[width=0.4\linewidth]{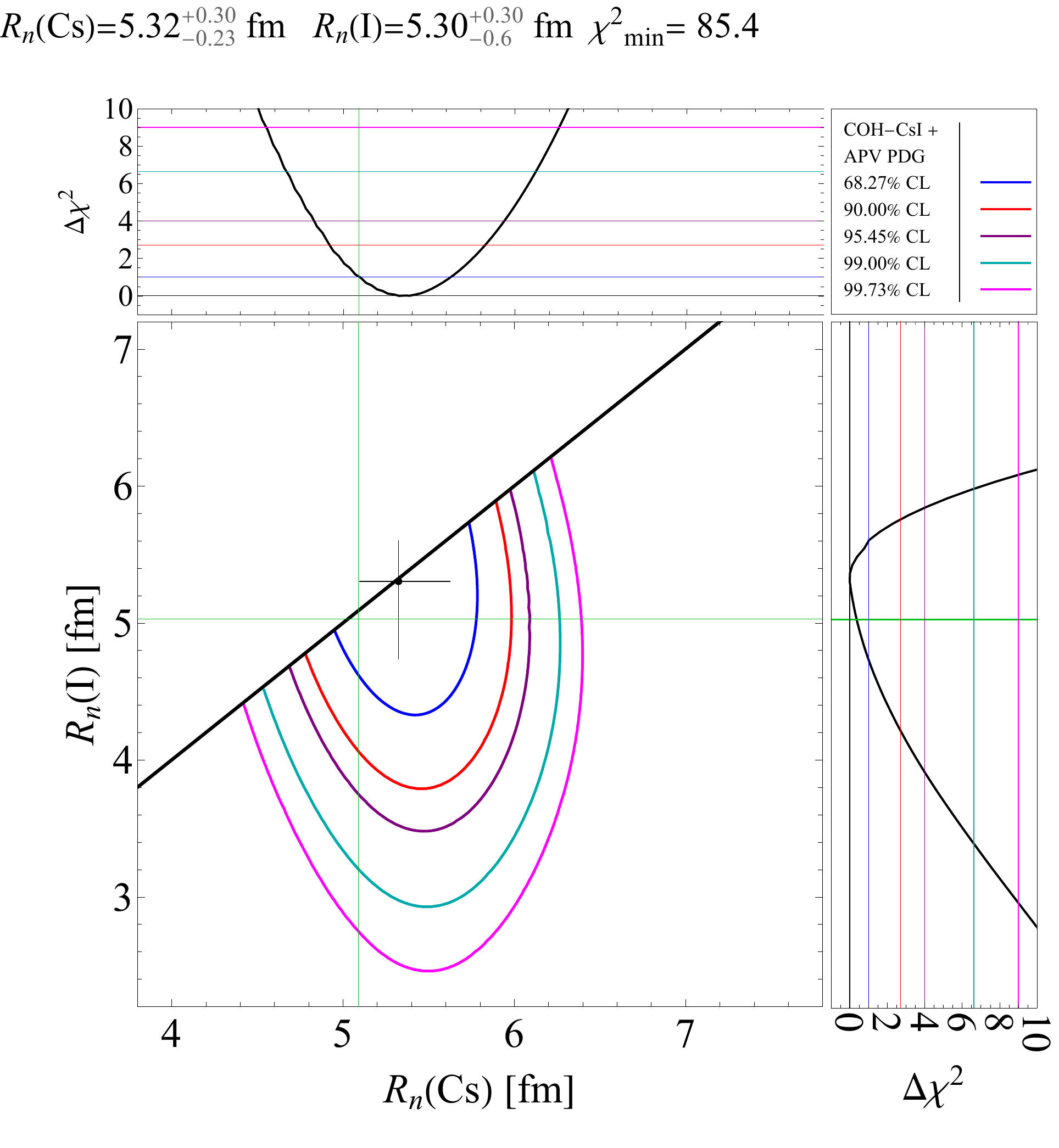}
\\
\end{tabular}
}
\end{tabular}
\begin{tabular}{cc}
\subfigure[]{\label{fig:2DRnCsRnIbCons}
\begin{tabular}{c}
\includegraphics*[width=0.4\linewidth]{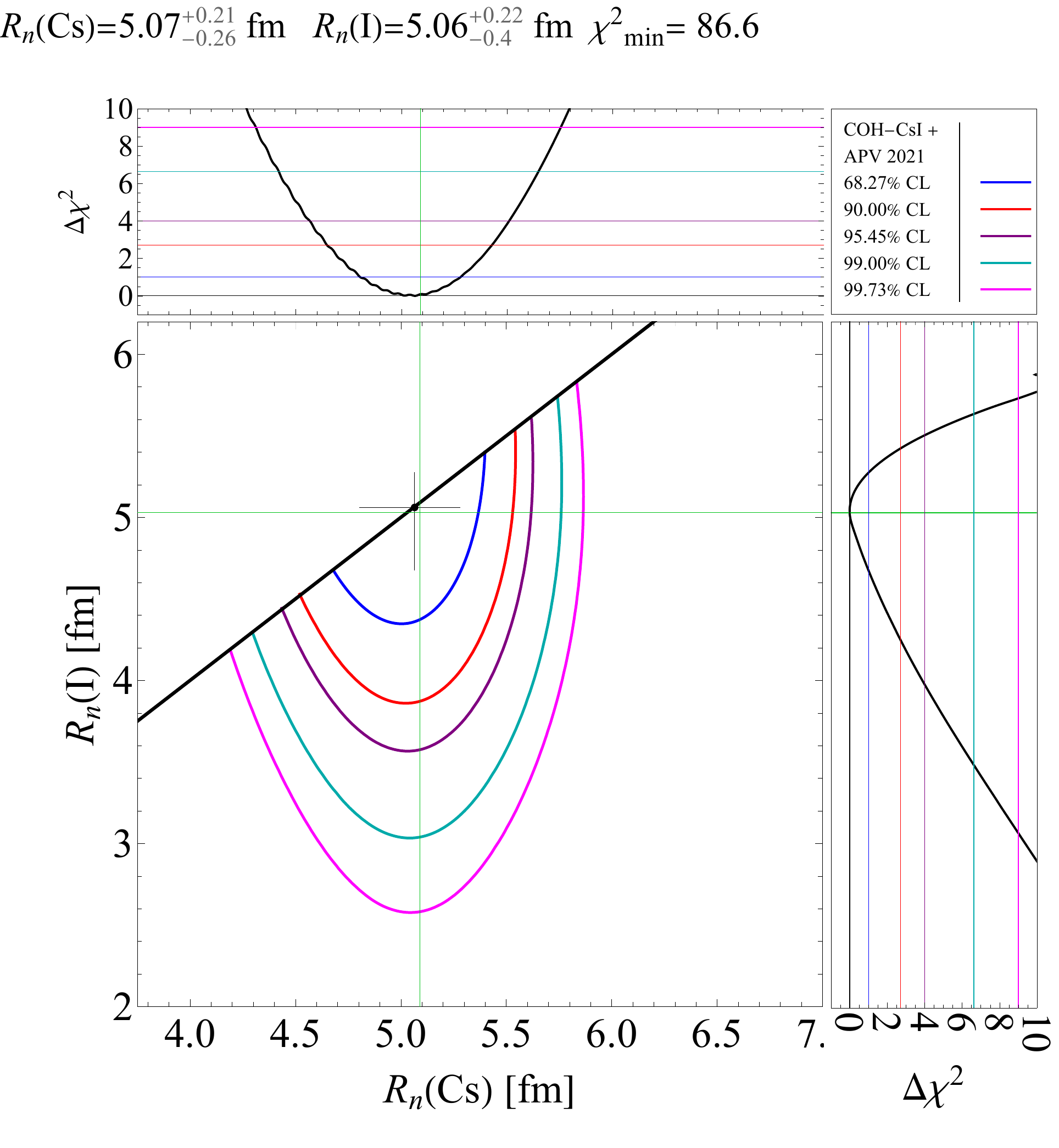}
\\
\end{tabular}
}
\\
\end{tabular}
\caption{ \label{fig:2DRnCsRnI}
Constraints on the plane of $R_n(^{133}{\rm{Cs}})$ and $R_n(^{127}{\rm{I}})$ together with their marginalizations, at different CLs obtained fitting the COHERENT CsI data alone (a) and in combination with APV data (b) and (c), using the value for the neutron skin corrections of Ref.~\cite{Dzuba:2012kx} (b) and Ref.~\cite{Sahoo:2021thl} (c). In all cases, we impose the constraint $R_n({\rm{Cs}})\geq R_n({\rm{I}})$.  The green lines indicate the corresponding NSM prediction for the average rms neutron radius of Cs and I.
}
\end{figure*}

In this appendix, we list the constraints on the plane of $R_n(^{133}{\rm{Cs}})$ and $R_n(^{127}{\rm{I}})$ together with their marginalizations, at different CLs, obtained fitting the COHERENT CsI data alone, Fig.~\ref{fig:RnCsRnICons}, and in combination with APV data, Figs.~\ref{fig:2DRnCsRnIaCons} and \ref{fig:2DRnCsRnIbCons}, imposing the constraint $R_n({\rm{Cs}})\geq R_n({\rm{I}})$. The results in combination with APV have been obtained using the value for the neutron skin correction of Ref.~\cite{Dzuba:2012kx} and the value recently calculated in Ref.~\cite{Sahoo:2021thl}, respectively.

\bibliography{ref}

\end{document}